 \documentclass[10pt,pra,aps,a4paper,oneside,twocolumn,superscriptaddress,showpacs]{revtex4-1}

\usepackage[english]{babel}
\usepackage[latin1]{inputenc}
\usepackage[T1]{fontenc}
\usepackage{lmodern}
\usepackage{ulem}
\usepackage{dcolumn}
\usepackage{subfigure}
\usepackage{footnote}
\usepackage{multirow}
\usepackage{amsmath,amsfonts,amssymb}

\usepackage{graphicx}  
\usepackage{latexsym}   
\usepackage{color}

\begin{document}

\title{Controlling photon bunching and antibunching of two quantum emitters near a core-shell sphere}

\author{\firstname{Tiago}  J. \surname{Arruda}}
\email{tiagojarruda@gmail.com}
\affiliation{Instituto de Ci\^encias Exatas,
Universidade Federal de Alfenas, 37133-840 Alfenas, Minas Gerais, Brazil}
\affiliation{Instituto de F\'isica de S\~ao Carlos,
Universidade de S\~ao Paulo, 13566-590 S\~ao Carlos, S\~ao Paulo, Brazil}

\author{\firstname{Romain}  \surname{Bachelard}}
\affiliation{Departamento de F\'isica,
Universidade Federal de S\~ao Carlos, 13565-905 S\~ao Carlos, S\~ao Paulo, Brazil}

\author{\firstname{John}  \surname{Weiner}}
\affiliation{Instituto de F\'isica de S\~ao Carlos,
Universidade de S\~ao Paulo, 13566-590 S\~ao Carlos, S\~ao Paulo, Brazil}

\author{\firstname{Sebastian}  \surname{Slama}}
\affiliation{Physikalisches Institut,
Eberhardt-Karls-Universit\"{a}t T\"{u}bingen, D-72076 T\"{u}bingen, Germany}

\author{\firstname{Philippe}  W. \surname{Courteille}}
\affiliation{Instituto de F\'isica de S\~ao Carlos,
Universidade de S\~ao Paulo, 13566-590 S\~ao Carlos, S\~ao Paulo, Brazil}

\begin{abstract}

    The collective spontaneous emission of two point-dipole emitters near a plasmonic core-shell nanosphere is theoretically investigated.
    Based on the expansion of mode functions in vector spherical harmonics, we derive closed analytical expressions for both the cooperative decay rate and the dipole-dipole interaction strength associated with two point dipoles close to a sphere.
    Considering a plasmonic nanoshell containing a linearly amplifying medium inside the core, the second-order correlation function for the two emitters shows that it is possible to tune the photon emission, selecting either photon bunching or antibunching as a function of the polarization and position of the sphere.
    This result opens vistas to applications involving tunable single-photon sources in engineered artificial media.

\end{abstract}


\maketitle


\section{Introduction}

Collective effects in the resonant fluorescence emitted by a system of many two-level atoms have been a subject of extensive research, both in theory and  practice.
The cooperative behavior of emitters at subwavelength scale can strongly modify the properties of the emitted radiation, leading, e.g., to the superradiance effect~\cite{Dicke_PhysRev93_1954,Lehmberg_PhysRevA2_1970,Ficek_Book_2005}.
From a theoretical perspective, rigorous analytical solutions for a many-atom system can only be obtained for systems of two or three atoms~\cite{Mvroyannis_PhysRevA18_1978,Agarwal_PhysRevA21_1980,Wiegand,Ficek_PhysRep372_2002}.
Indeed, the general many-atom case requires some approximations to decouple the dynamical equations and make them suitable for numerical calculations~\cite{Jenkins_PhysRevA94_2016}.
Due to its simplicity, the two-atom system has been used as a prototype model for studying collective phenomena and hence isolating the single-atom effects from those arising from correlations in many-atom systems~\cite{Ficek_PhysRep372_2002}.
With the recent technical advances in confining a few atoms or ions at small interatomic separations~\cite{Brewer_PhysRevLett76_1996}, arranged into a linear chain~\cite{Ficek_PhysA146_1987,Scully_PhysRevLett101_2008,Blatt_Nature413_2001,Cirac_PhysRevA78_2008,Zubairy_PhysRevA90_2014,Sandoghdar_Science298_2002,Zubairy_PhysRevA73_2006} or two-dimensional arrays~\cite{Faridani_PhysRevLett105_1984,Freedhoff_PhysRevA69_1986}, there has been a renewal of interest in the study of systems consisting of two optical emitters for technological applications.

In addition to collective effects, the resonant emission of light by each emitter can also be modified by the interaction between single emitters and the electromagnetic environment, which is generally referred to as the Purcell effect~\cite{Purcell_PhysRev69_1946,Chew_JCPhys87_1987,Dereux_PhysRevB84_2011,Bordo_JOSAB31_2014,Belov_SciRep5_2015}.
Owing to their ability to concentrate light at subwavelength scales, a great deal of attention has been devoted to manipulating light emission and absorption via the Purcell effect using plasmonic nanostructures~\cite{Moroz_ChemPhys317_2005,Farina_PhysRevA87_2013,Liu_Nature9_2014,Carminati_SurfSciRep70_2015,Belov_SciRep5_2015,Szilard_PhysRevB94_2016,Cuevas_JOpt18_2016,Girard_JOpt18_2016,Arruda_PhysRevA96_2017,Arruda_Springer219_2018,Arruda_PhysRevB98_2018}.
In these systems, a strong Purcell effect associated with the enhancement or suppression of the spontaneous emission is achieved due to the excitation of surface plasmons on metal-insulator interfaces, which modifies the local density of states (LDOS) and hence the spontaneous decay rates~\cite{Lukin_PhysRevLett97_2006,Lukin_Nature450_2007,
Shahbazyan_PhysRevLett102_2009,Shahbazyan_PhysRevLett117_2016,Sun_SciApp4_2015,Girard_JOpt18_2016,Gu_SciRep8_2018}.
More recently, the collective coupling between quantum emitters and localized surface plasmons has been realized in a plasmonic system~\cite{Vidal_PhysRevLett112_2014,Slama_NatPhys10_2014}.

In general, all applications involving plasmonic materials are limited by high Ohmic losses on metallic surfaces~\cite{Carminati_SurfSciRep70_2015,Shahbazyan_PhysRevLett117_2016}.
In fact, in the case of spontaneous emission in the vicinity of a plasmonic structure, the radiative and nonradiative contributions to the Purcell factor must be clearly identified since non-radiative channels are dominant in the near field~\cite{Arruda_PhysRevA96_2017,Arruda_Springer219_2018}.
Among different proposals to minimize losses in plasmonic structures, one solution is to include a gain material within the system in order to compensate losses via stimulated emission of plasmons~\cite{Shalaev_Nature466_2010,Tsakmakidis_Science339_2013}.
Indeed, the use of loss compensation in plasmonic nanoshells to achieve composite metamaterials with near-zero permittivity has been recently proposed~\cite{Capolino_OptMatExp1_2011,Capolino_Nanotech23_2012}.
In addition, a gain-assisted plasmonic sphere may exhibit very narrow visible higher-order modes which are usually dominated in the spectrum by the broad spectral features of lower-order modes~\cite{Klar_Nano4_2013}.

In this paper, we propose the application of gain-assisted plasmonic nanoshells to enhance, tailor, and control correlations in the fluorescence emitted by a system of two point-dipole emitters.
The main idea is to achieve a system that can exhibit both photon-bunching and antibunching effects~\cite{Marty_PhysRevB82_2010,Elmer_NewJPhys21_2019} depending on the polarization of the incident laser field and the gain.
The properties of the emitted radiation depend on several parameters that describe the dipole emitters, the geometry of the system, the electromagnetic environment, and the laser field.
All of these parameters are encoded in correlation functions of the emitted field amplitude.
Here, we use the two-time second-order correlation function $g^{(2)}(\tau)$, with $\tau=t_2-t_1$, which provides these emission properties of the radiated field determining whether it is classical or quantum in nature~\cite{Ficek_PhysRep372_2002}.
In particular, we consider in our investigation a well-known expression for the second-order correlation function obtained by Wiegand~\cite{Wiegand} for two interacting and weakly driven two-level atoms in equivalent positions.

To calculate the collective decay rate and the frequency shifts to enter into the correlation functions, we generalize previous studies on the spontaneous emission of single-dipole emitters close to a plasmonic sphere~\cite{Arruda_PhysRevA96_2017,Arruda_Springer219_2018,Arruda_PhysRevB98_2018} to the case of a two-atom system~\cite{Ficek_PhysRep372_2002}.
Here, our analytical expressions concerning the decay rates and frequency shifts are valid for two arbitrary point-dipole emitters (e.g., two-level atoms, ions, molecules or quantum dots) in the vicinity of a spherical body.
These analytical expressions for the spherical case are important to benchmark numerical calculations in order to characterize more complex geometries.

The remainder of this paper is organized as follows.
In Sec.~\ref{Spontaneous}, we derive closed analytical expressions for the spontaneous emission rate and the dipole-dipole interaction strength of two dipole emitters in the vicinity of a sphere.
These are the main analytical results of our study.
A brief review of the two-time second-order correlation function for a two-atom system is provided in Sec.~\ref{Second-time}.
The scattering properties of a plasmonic core-shell sphere within the framework of the Lorenz-Mie theory are presented in Sec.~\ref{sec-scattering}.
Section~\ref{Gain-assisted} is devoted to collective effects of two atoms near a plasmonic core-shell sphere containing a gain material.
Finally, in Sec.~\ref{Conclusion}, we summarize our results and conclude.

\section{Spontaneous emission of two two-level atoms in close proximity to a sphere}
\label{Spontaneous}

Let us consider two two-level atoms, located at $\mathbf{r}_1$ and $\mathbf{r}_2$, that can be well described by two of their eigenstates, $\{|{\rm g}_q\rangle,|{\rm e}_q\rangle\}$, where $q=1$ for atom 1 and $q=2$ for atom 2.
As usual, $|{\rm g}_q\rangle$ is the eigenstate with lowest energy ($E_{{\rm g}_q}=-\hbar\omega_q/2$), i.e., the ground state, whereas $|{\rm e}_q\rangle$ is the eigenstate with highest energy ($E_{{\rm e}_q}=\hbar\omega_q/2$) coupled to $|{\rm g}_q\rangle$ by an electric dipole moment $\mathbf{d}_{q}\equiv\langle{\rm g}_q|\hat{\mathbf{d}}_q|{\rm e}_q\rangle$~\cite{Milonni_Book1994}.
Details of the corresponding Hamiltonian of the two-atom system in the electric dipole approximation are given in Appendix~\ref{Hamiltonian}.

By solving the Heisenberg equations of motion for the atomic and field operators within the Born and Markov approximations, one obtains the spontaneous emission rate on a transition $|{\rm e}_q\rangle \to |{\rm g}_q\rangle$ of frequency $\omega_q$: $\gamma_{q}^{\rm rad}=2\pi\sum_{\mathbf{k}p}|g_{\mathbf{k}p}(\mathbf{r}_q)|^2\delta(\omega_{\mathbf{k}}-\omega_q)$, where $g_{\mathbf{k}p}(\mathbf{r}_q)$ is the atom-field coupling coefficient associated with atom $q$ and a field with wave vector $\mathbf{k}$ and polarization $p$~\cite{Milonni_Book1994}.
In terms of the mode functions $\mathbf{u}_{\mathbf{k}p}(\mathbf{r}_q)$ defined in Appendix~\ref{Hamiltonian}, which are solutions of a vector Helmholtz equation related to the electromagnetic environment, one has
\begin{align}
\gamma_{q}^{\rm rad}(\mathbf{r}_{q})=\frac{\pi\omega_q}{\varepsilon_0\hbar}\sum_{\mathbf{k}p}\left|\mathbf{d}_q\cdot\mathbf{u}_{\mathbf{k}p}(\mathbf{r}_q)\right|^2\delta(\omega_{\mathbf{k}}-\omega_q),\label{Gamma-single}
\end{align}
which is the same result obtained by the Weisskopf-Wigner theory~\cite{Milonni_Book1994}.
In vacuum, Eq.~(\ref{Gamma-single}) retrieves the well-known result for the Einstein $A$ coefficient:
\begin{align}
\gamma_{q}^{(0)}\equiv\frac{|\mathbf{d}_q|^2\omega_q^3}{3\pi\varepsilon_0\hbar c^3}.
\end{align}

In addition, due to the coupling between the atoms through the vacuum field, one also has the cross-damping spontaneous emission rate~\cite{Ficek_PhysA146_1987,Akram_PhysRevA62_2000}: $\gamma_{12}^{\rm rad}=\gamma_{21}^{\rm rad}=2\pi\sum_{\mathbf{k}p}{\rm Re}[g_{\mathbf{k}p}(\mathbf{r}_1)g_{\mathbf{k}p}^*(\mathbf{r}_2)]\delta(\omega_{\mathbf{k}}-\omega_0)$, where $\omega_0=(\omega_1+\omega_2)/2$.
Using Eq.~(\ref{g-coupling}), we obtain the cross-damping decay rate

\begin{align}
&\gamma_{12}^{\rm rad}(\mathbf{r}_1,\mathbf{r}_2)=\gamma_{21}^{\rm rad}(\mathbf{r}_2,\mathbf{r}_1)\nonumber\\
&=\frac{\pi\omega_0}{\varepsilon_0\hbar}\sum_{\mathbf{k}p}{{\rm Re}\left[\mathbf{d}_1\cdot\mathbf{u}_{\mathbf{k}p}(\mathbf{r}_1)\mathbf{u}_{\mathbf{k}p}^*(\mathbf{r}_2)\cdot\mathbf{d}_2^*\right]}\delta(\omega_{\mathbf{k}}-\omega_0),\label{Gamma-coupling}
\end{align}
which shows explicitly the cooperative effect of the dipole-dipole interaction in the spontaneous emission rate.

\subsection{Radiative decay rates of two excited atoms near a sphere}

The analytical expressions for the spontaneous decay rates associated with a single quantum emitter in the vicinity of a sphere are well-known~\cite{Ruppin_JCPhys76_1982,Chew_JCPhys87_1987}.
They can be readily calculated by substituting the vector mode functions from Appendix~\ref{Mode-functions}, Eqs.~(\ref{Akx-inc})-(\ref{Aky-sca}), into Eq.~(\ref{Gamma-single}):
 \begin{align}
\gamma_q^{\rm rad}(k_qr_q) &= |\hat{\mathbf{d}}_q\cdot\hat{\mathbf{r}}|^2\gamma_{q\perp}^{\rm rad}(k_qr_q)\nonumber\\
 &+ (1-|\hat{\mathbf{d}}_q\cdot\hat{\mathbf{r}}|^2)\gamma_{q||}^{\rm rad}(k_qr_q),\label{eq-agora}
\end{align}
where the contributions of the electric dipole moment $\mathbf{d}_q$ oriented orthogonal $(\perp)$ or parallel $(||)$ to the spherical surface are, respectively,
\begin{align}
\frac{\gamma_{q\perp}^{\rm rad}(k_qr_q)}{\gamma_q^{(0)}}&=\frac{3}{2}\sum_{\ell=1}^{\infty} \ell(\ell+1)(2\ell+1)\nonumber\\
&\times\left|\frac{j_{\ell}(k_qr_q)-a_{\ell}h_{\ell}^{(1)}(k_qr_q)}{k_qr_q}\right|^2,\label{Gamma-perp-rad}\\
\frac{\gamma_{q||}^{\rm rad}(k_qr_q)}{\gamma_q^{(0)}}&=\frac{3}{4}\sum_{\ell=1}^{\infty}(2\ell+1)\Bigg[\left|\frac{\psi_{\ell}'(k_qr_q)-a_{\ell}\xi_{\ell}'(k_qr_q)}{k_qr_q}\right|^2\nonumber\\
&+\left|j_{\ell}(k_qr_q)-b_{\ell}h_{\ell}^{(1)}(k_qr_q)\right|^2\Bigg],\label{Gamma-para-rad}
\end{align}
with $q=\{1,2\}$ and $k_q=\omega_q/c$.
If the atomic dipoles have arbitrary orientations in relation to the spherical surface, one can consider the spatial mean: $|\hat{\mathbf{d}}_q\cdot\hat{\mathbf{r}}|^2=1/3$.
To derive Eqs.~(\ref{Gamma-perp-rad}) and (\ref{Gamma-para-rad}) we have used the mode functions $\mathbf{u}_{\mathbf{k}p}(\mathbf{r}_q)$ defined in Appendix~\ref{Mode-functions} and the relations~\cite{Bohren_Book_1983}: $\int_{-1}^1{\rm d}(\cos\theta)(\pi_{\ell}\pi_{\ell'}+\tau_{\ell}\tau_{\ell'})=[2\ell^2(\ell+1)^2/(2\ell+1)]\delta_{\ell\ell'}$, $\int_{-1}^1{\rm d}(\cos\theta)(\pi_{\ell}\tau_{\ell'}+\tau_{\ell}\pi_{\ell'})=0$, and $\int_{-1}^1{\rm d}(\cos\theta)\sin^2\theta\pi_{\ell}\pi_{\ell'}=[2\ell(\ell+1)/(2\ell+1)]\delta_{\ell\ell'}$, where $\pi_{\ell}$ and $\tau_{\ell}$ are generalized Legendre functions.
The corresponding decay rates in free space are retrieved by taking $a_{\ell}=0=b_{\ell}$ and using the identities~\cite{Chew_JCPhys87_1987}
\begin{align}
&\sum_{\ell=1}^{\infty}\ell(\ell+1)(2\ell+1)\frac{j_{\ell}^2(k_qr_q)}{(k_qr_q)^2}=\frac{2}{3},\label{Bessel1}\\
&\sum_{\ell=1}^{\infty}(2\ell+1)\left[j_{\ell}^2(k_qr_q)+\frac{\psi_{\ell}'^2(k_qr_q)}{(k_0r_q)^2}\right]=\frac{4}{3},\label{Bessel2}
\end{align}
and hence $\gamma_{q\perp}^{\rm rad}=\gamma_{q||}^{\rm rad}=\gamma_q^{(0)}$.
Using these general ideas, before considering the influence of a spherical body on two excited atoms, it is convenient to determine the cooperative decay rate of two atoms in free space, and then generalize it to the case of a spherical body in their vicinity.

\begin{figure}[htbp]
\centerline{\includegraphics[width=\columnwidth]{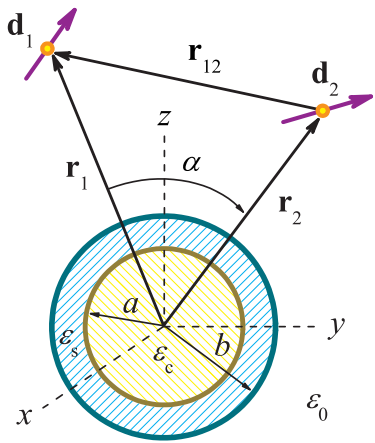}}
\caption{Two point-dipole emitters located in the vicinity of a coated sphere.
The core-shell sphere has inner radius $a$ and outer radius $b$, with electric permittivities $\varepsilon_{\rm c}$ for the core and $\varepsilon_{\rm s}$ for the shell.
The dipole emitters are characterized by the atomic dipole moment $\mathbf{d}_q$ and are located at $\mathbf{r}_q$, with $q=\{1,2\}$.
The interatomic distance is $r_{12}=|\mathbf{r}_{12}|=|\mathbf{r}_1-\mathbf{r}_2|$, with $\alpha$ being the angle between $\mathbf{r}_1$ and $\mathbf{r}_2$.}\label{fig1}
\end{figure}

For two two-level atoms in free space, we can use directly the mode functions $\mathbf{u}_{\mathbf{k}x}(\mathbf{r})=e^{\imath kr\cos\theta_k}(-\sin\theta_k\mathbf{e}_k + \cos\theta_k\mathbf{e}_{\theta_k})/\sqrt{V}$, and
$\mathbf{u}_{\mathbf{k}y}(\mathbf{r})=e^{\imath kr\cos\theta_k}\mathbf{e}_{\varphi_k}/\sqrt{V}$ ($V$ the quantization volume) instead of Eqs.~(\ref{Akx-inc}) and (\ref{Aky-inc}), which are the corresponding expansions in spherical wave functions.
Here, we have no restriction on the coordinate system to perform the integrals on $k$-space.
This is due to the fact that the spontaneous decay rates in free space depend only on the interatomic distance $r_{12}=|\mathbf{r}_1-\mathbf{r}_2|$ instead of $\mathbf{r}_1$ and $\mathbf{r}_2$.
Without loss of generality, we choose a coordinate system in which $\mathbf{r}_1$ is parallel to $\mathbf{r}_2$, which implies $|\mathbf{r}_1-\mathbf{r}_2|=|r_1-r_2|$, i.e., $\mathbf{k}\cdot(\mathbf{r}_1-\mathbf{r}_2)=k(r_1-r_2)\cos\theta_k$.
Using the definition in Eq.~(\ref{Gamma-coupling}), for the radial and non-radial orientations of the atomic dipole (in relation to the chosen spherical coordinate system), we obtain the well known cross-damping decay rates in free space~\cite{Agarwal_PhysRevA12_1975,Ficek_PhysRep372_2002}:
\begin{align}
\frac{\gamma_{12\perp}^{(0)}}{\sqrt{\gamma_1^{(0)}\gamma_2^{(0)}}}&=3\left[-\frac{\cos(k_0r_{12})}{(k_0r_{12})^2}+\frac{\sin(k_0r_{12})}{(k_0r_{12})^3}\right],\label{Gamma12-perp-vac}\\
\frac{\gamma_{12||}^{(0)}}{\sqrt{\gamma_1^{(0)}\gamma_2^{(0)}}}&=\frac{3}{2}\left[\frac{\sin(k_0r_{12})}{k_0r_{12}} +\frac{\cos(k_0r_{12})}{(k_0r_{12})^2}- \frac{\sin(k_0r_{12})}{(k_0r_{12})^3}\right],\label{Gamma12-para-vac}
\end{align}
where we have considered $r_{12}=|\mathbf{r}_1-\mathbf{r}_2|$ and \color{black} $k_0=\omega_0/c$, with $\omega_0=(\omega_1+\omega_2)/2$ and $|\omega_1-\omega_2|\ll\omega_0$.
In addition, since $\mathbf{r}_1$ and $\mathbf{r}_2$ are parallel, we can introduce the total cross-damping decay rate in free space as
\begin{align}
\gamma_{12}^{(0)}&={\rm Re}\bigg\{(\hat{\mathbf{d}}_1\cdot\hat{\mathbf{r}}_{12})(\hat{\mathbf{d}}_2^*\cdot\hat{\mathbf{r}}_{12})\gamma_{12\perp}^{(0)}\nonumber\\
 &+ \left[\hat{\mathbf{d}}_1\cdot\hat{\mathbf{d}}_2^*-(\hat{\mathbf{d}}_1\cdot\hat{\mathbf{r}}_{12})(\hat{\mathbf{d}}_2^*\cdot\hat{\mathbf{r}}_{12})\right]\gamma_{12||}^{(0)}\bigg\},\label{gamma12-welll}
\end{align}
where the radial and non-radial contributions are given in Eqs.~(\ref{Gamma12-perp-vac}) and (\ref{Gamma12-para-vac}), respectively.
Substituting Eqs.~(\ref{Gamma12-perp-vac}) and (\ref{Gamma12-para-vac}) into Eq.~(\ref{gamma12-welll}), one retrieves the well-known expression for the cross-damping decay rate in free space~\cite{Agarwal_PhysRevA45_1992,Akram_PhysRevA62_2000}.
Note that Eq.~(\ref{gamma12-welll}) was here introduced as a generalization of the single-emitter case, Eq.~(\ref{eq-agora}), to the case of two emitters by imposing $\mathbf{r}_{12}$ parallel to $\mathbf{r}$.
\color{black}

Of course, one could also calculate the above cross-damping decay rates in free space using the expansions of plane waves in terms of spherical harmonics, Eqs.~(\ref{Akx-inc}) and (\ref{Aky-inc}) in Appendix~\ref{Mode-functions}.
Since there is no restriction on the coordinate system, one could again consider that $\mathbf{r}_1$ and $\mathbf{r}_2$ are parallel.
However, within this framework, we can explore the properties of spherical functions to obtain a general result for any spherical coordinate system, in which there is an arbitrary angle $\alpha$ between $\mathbf{r}_1$ and $\mathbf{r}_2$, see Fig.~\ref{fig1}:
\begin{align}
r_{12}^2 = r_1^2 + r_2^2 - 2r_1r_2\cos\alpha.\label{r12}
\end{align}
To this end, we use the addition theorem of spherical harmonics, which can be reduced to the following expression for the spherical Bessel function~\cite{Abramovitz_book_1964}:
\begin{align}
&\sum_{\ell=0}^{\infty}(2\ell+1)j_{\ell}(k_0r_1)j_{\ell}(k_0r_2)P_{\ell}(\cos\alpha)=\frac{\sin(k_0r_{12})}{k_0r_{12}},\label{Bessel3}
\end{align}
where $P_{\ell}(\cos\alpha)$ is a Legendre polynomial.
By deriving Eq.~({\ref{Bessel3}) in relation to $\cos\alpha$ and recalling the definition $P_{\ell}'(\cos\alpha)=\pi_{\ell}(\cos\alpha)$~\cite{Bohren_Book_1983}, we obtain
\begin{align}
&\sum_{\ell=1}^{\infty} (2\ell+1)\frac{j_{\ell}(k_0r_1)}{k_0r_1}\frac{j_{\ell}(k_0r_2)}{k_0r_2}\pi_{\ell}(\cos\alpha)\nonumber\\
&=-\frac{\cos(k_0r_{12})}{(k_0r_{12})^2}+\frac{\sin(k_0r_{12})}{(k_0r_{12})^3}.\label{Bessel5}
\end{align}
Since $\pi_{\ell}(1)=\ell(\ell+1)/2$, as $\alpha=0$ and ${r}_1\to{r}_2$, Eq.~(\ref{Bessel5}) is equivalent to Eq.~(\ref{Bessel1}) for $k_q=k_0$.
Although it is not straightforward, a generalization of Eq.~(\ref{Bessel2}) can also be obtained.
Indeed, we have verified that
\begin{align}
&\sum_{\ell=1}^{\infty}\frac{(2\ell+1)}{\ell(\ell+1)}\bigg[j_{\ell}(k_0r_1)j_{\ell}(k_0r_2)\tau_{\ell}(\cos\alpha)\nonumber\\
&+\frac{\psi_{\ell}'(k_0r_1)}{k_0r_1}\frac{\psi_{\ell}'(k_0r_2)}{k_0r_2}\pi_{\ell}(\cos\alpha)\bigg]\nonumber\\
&=\frac{\sin(k_0r_{12})}{k_0r_{12}}+\frac{\cos(k_0r_{12})}{(k_0r_{12})^2}-\frac{\sin(k_0r_{12})}{(k_0r_{12})^3},\label{Bessel6}
\end{align}
which corresponds to Eq.~(\ref{Bessel2}) for $\alpha=0$ and ${r}_1\to{r}_2$.

Provided any arbitrary spherical coordinate system, Eqs.~(\ref{Bessel5}) and (\ref{Bessel6}) allow us to calculate the cooperative decay rate of two dipole emitters in free space in terms of spherical wave functions, which is one of the main analytical results of our study.
The general expression of the cooperative decay rate in terms of Bessel functions is readily obtained by substituting Eqs.~(\ref{Bessel5}) and (\ref{Bessel6}) into Eqs.~(\ref{Gamma12-perp-vac}) and (\ref{Gamma12-para-vac}), respectively.
\color{black}
Note that Eq.~(\ref{gamma12-welll}) remains unchanged for $\alpha\not=0$.

Now, we can finally calculate the cross-damping decay rate of two quantum emitters near a sphere centered at the origin of a spherical coordinate system.
To be consistent with the well-known result in free space, Eq.~(\ref{gamma12-welll}), the radiative contribution of the total cross-damping decay rate must have the form
\color{black}
\begin{align}
\gamma_{12}^{\rm rad}&={\rm Re}\bigg\{(\hat{\mathbf{d}}_1\cdot\hat{\mathbf{r}}_{12})(\hat{\mathbf{d}}_2^*\cdot\hat{\mathbf{r}}_{12})\gamma_{12\perp}^{\rm rad}\nonumber\\
 &+ \left[\hat{\mathbf{d}}_1\cdot\hat{\mathbf{d}}_2^*-(\hat{\mathbf{d}}_1\cdot\hat{\mathbf{r}}_{12})(\hat{\mathbf{d}}_2^*\cdot\hat{\mathbf{r}}_{12})\right]\gamma_{12||}^{\rm rad}\bigg\},\label{gamma12-well}
\end{align}
where $\gamma_{12\perp}^{\rm rad}$ and $\gamma_{12||}^{\rm rad}$ are calculated by using Eq.~(\ref{Gamma-coupling}) and the corresponding mode functions, Eqs.~(\ref{Akx-inc})--(\ref{Aky-sca}) in Appendix~\ref{Mode-functions}.
After some calculations, we obtain
\begin{widetext}
\begin{align}
\frac{\gamma_{12\perp}^{\rm rad}(k_0r_1,k_0r_2,\cos\alpha)}{\sqrt{\gamma_1^{(0)}\gamma_2^{(0)}}}&={3}\sum_{\ell=1}^{\infty} (2\ell+1){\rm Re}\Bigg\{\left[\frac{j_{\ell}(k_0r_1)-a_{\ell}h_{\ell}^{(1)}(k_0r_1)}{k_0r_1}\right]\left[\frac{j_{\ell}(k_0r_2)-a_{\ell}^*h_{\ell}^{(1)*}(k_0r_2)}{k_0r_2}\right]\Bigg\}\pi_{\ell}(\cos\alpha),\label{Gamma12-perp-rad}\\
\frac{\gamma_{12||}^{\rm rad}(k_0r_1,k_0r_2,\cos\alpha)}{\sqrt{\gamma_1^{(0)}\gamma_2^{(0)}}}&=\frac{3}{2}\sum_{\ell=1}^{\infty}\frac{(2\ell+1)}{\ell(\ell+1)}{\rm Re}\Bigg\{\left[\frac{\psi_{\ell}'(k_0r_1)-a_{\ell}\xi_{\ell}'(k_0r_1)}{k_0r_1}\right]\left[\frac{\psi_{\ell}'(k_0r_2)-a_{\ell}^*\xi_{\ell}'^*(k_0r_2)}{k_0r_2}\right]\pi_{\ell}(\cos\alpha)\nonumber\\
&+\left[j_{\ell}(k_0r_1)-b_{\ell}h_{\ell}^{(1)}(k_0r_1)\right]\left[j_{\ell}(k_0r_2)-b_{\ell}^*h_{\ell}^{(1)*}(k_0r_2)\right]\tau_{\ell}(\cos\alpha)\Bigg\},\label{Gamma12-para-rad}
\end{align}
\end{widetext}
\color{black}
where we have used Eqs.~(\ref{Bessel5}) and (\ref{Bessel6}) to generalize the case $\alpha=0$ to $0\leq\alpha\leq\pi$.
Due to this generalization, note in Eq.~(\ref{gamma12-well}) that the subscripts indicating the contributions of dipole moments $\mathbf{d}_q$ oriented orthogonal ($\perp$) or parallel ($||$) to the particle surface have a straightforward interpretation only for $\alpha=0$ or $\alpha=\pi$, i.e., when $\mathbf{r}_{12}$ is parallel to $\mathbf{r}$.
Indeed, for $\alpha\not=0$ and $\alpha\not=\pi$, the prefactor of $\gamma_{12\perp}^{\rm rad}$ in Eq.~(\ref{gamma12-well}) indicates the projection of $\mathbf{d}_q$ onto $\mathbf{r}_{12}$, with $q=\{1,2\}$.
Here we consider that the direction of the electric dipole moments $\mathbf{d}_q$ coincides with the direction of the local electric field outside the sphere at the point of space the emitter is located.
This electric field can be calculated using the Lorenz-Mie theory given in Appendix~\ref{Lorenz-Mie}.
\color{black}

\subsection{Nonradiative decay rates near a sphere and frequency shifts}

The theory provided so far only tells us how to calculate the radiative contribution of the decay rates.
In order to include nonradiative contributions on the expressions calculated above, we apply a heuristic approach based on energy conservation in the Lorenz-Mie theory.
The Lorenz-Mie theory describes the light scattering by a sphere of arbitrary radius, where the scattering coefficients $a_{\ell}$ and $b_{\ell}$ are associated with the extinction, scattering and absorption cross sections, respectively~\cite{Bohren_Book_1983}:
\begin{align}
\sigma_{\rm ext} &= \frac{2\pi}{k^2}\sum_{\ell=1}^{\infty}(2\ell+1){\rm Re}\left(a_{\ell}+b_{\ell}\right),\label{Qext}\\
\sigma_{\rm sca} &= \frac{2\pi}{k^2}\sum_{\ell=1}^{\infty}(2\ell+1)\left(|a_{\ell}|^2+|b_{\ell}|^2\right),\\
\sigma_{\rm abs} &= \sigma_{\rm ext}-\sigma_{\rm sca}.\label{Qabs}
\end{align}
In the absence of absorption, $\sigma_{\rm abs}=0$ and one has $|a_{\ell}|^2={\rm Re}(a_{\ell})$ and $|b_{\ell}|^2={\rm Re}(b_{\ell})$, {i.e.}, the extinction and scattering cross sections are interchangeable: $\sigma_{\rm ext}=\sigma_{\rm sca}$.
This allows us to derive the total decay rate $\gamma$ from the radiative contributions by rewriting Eqs.~(\ref{Gamma-perp-rad})--(\ref{Gamma12-para-rad}) in terms of $|a_{\ell}|^2$ and $|b_{\ell}|^2$, and then changing $|a_{\ell}|^2\to{\rm Re}(a_{\ell})$ and $|b_{\ell}|^2\to{\rm Re}(b_{\ell})$.
From this procedure, we obtain well-known expressions for the total decay rates of a single quantum emitter at $\mathbf{r}_q$~\cite{Chew_JCPhys87_1987}:
\begin{align}
\frac{\gamma_{q\perp}(k_qr_q)}{\gamma_q^{(0)}}&=1-\frac{3}{2}\sum_{\ell=1}^{\infty}\ell(\ell+1)(2\ell+1)\nonumber\\
&\times{\rm Re}\left\{a_{\ell}\left[\frac{h_{\ell}^{(1)}(k_qr_q)}{k_qr_q}\right]^2\right\},\label{Gamma-perp}\\
\frac{\gamma_{q||}(k_qr_q)}{\gamma_q^{(0)}}&=1-\frac{3}{4}\sum_{\ell=1}^{\infty}(2\ell+1){\rm Re}\Bigg\{a_{\ell}\left[\frac{\xi_{\ell}'(k_qr_q)}{k_qr_q}\right]^2 \nonumber\\
&+ b_{\ell}\left[ h_{\ell}^{(1)}(k_qr_q)\right]^2\Bigg\}.\label{Gamma-para}
\end{align}

In addition, the Green's tensor formalism associates the real and imaginary parts of the Green function (dotted into the dipole moment) to the total decay rate $\gamma$ and the shift $\delta$ of the transition frequency due to the presence of the sphere, respectively~\cite{Welsch_PhysRevA64_2001}.
Applying this formalism to Eqs.~(\ref{Gamma-perp}) and (\ref{Gamma-para}), one has the corresponding frequency shifts~\cite{Letokhov_JModOpt43_2_1996}
\begin{align}
\frac{\delta_{q}^{\perp}(k_qr_q)}{\gamma_q^{(0)}}&=\frac{3}{4}\sum_{\ell=1}^{\infty}\ell(\ell+1)(2\ell+1){\rm Im}\left\{a_{\ell}\left[\frac{h_{\ell}^{(1)}(k_qr_q)}{k_qr_q}\right]^2\right\},\label{delta-perp}\\
\frac{\delta_{q}^{||}(k_qr_q)}{\gamma_q^{(0)}}&=\frac{3}{8}\sum_{\ell=1}^{\infty}(2\ell+1){\rm Im}\Bigg\{a_{\ell}\left[\frac{\xi_{\ell}'(k_qr_q)}{k_qr_q}\right]^2 \nonumber\\
&+ b_{\ell} \left[h_{\ell}^{(1)}(k_qr_q)\right]^2\Bigg\},\label{delta-para}
\end{align}
where $\delta_q = |\hat{\mathbf{d}}_q\cdot\hat{\mathbf{r}}|^2\delta_{q}^{\perp} + (1-|\hat{\mathbf{d}}_q\cdot\hat{\mathbf{r}}|^2)\delta_{q}^{||}$~\cite{Letokhov_JModOpt43_2_1996}.
Usually, the frequency shift $\delta_q$ is already encoded in $\omega_q$ in the calculations.
To simplify our discussion, we assume from now on that $\omega_q\equiv\omega_q+\delta_q$.

Similarly, for the case of two atoms, the cooperative decay rate $\gamma_{12}$ is associated with a frequency shift $\delta_{12}$ of the atomic levels, known as the retarded dipole-dipole interaction.
From a fully quantum treatment of two emitters, $\gamma_{12}$ and $\delta_{12}$ are related to each other through a response function $\chi_{\alpha\beta}(\mathbf{r}_1,\mathbf{r}_2,\omega_0)$~\cite{Agarwal_PhysRevA45_1992,Agarwal_PhysRevA57_1998}:
\begin{align}
2\delta_{12}-\imath\gamma_{12}=-\frac{1}{\hbar}\sum_{\alpha,\beta}d_1^{\alpha}d_2^{\beta}\chi_{\alpha\beta}(\mathbf{r}_1,\mathbf{r}_2,\omega_0).\label{Chi}
\end{align}
The response function can be understood as the $\alpha$th component of the electric field at $\mathbf{r}_1$ produced by an oscillating dipole at $\mathbf{r}_2$ and oriented along the $\beta$ direction.
Since the response function in free space $\chi_{\alpha\beta}(\mathbf{r}_1,\mathbf{r}_2,\omega_0)\propto e^{\imath k_0 r_{12}}/k_0r_{12}$, we readily obtain the corresponding retarded dipole-dipole interaction from Eqs.~(\ref{Gamma12-perp-vac}), (\ref{Gamma12-para-vac}), and (\ref{Chi})~\cite{Agarwal_PhysRevA12_1975,Ficek_PhysRep372_2002}:
\begin{align}
\frac{\delta_{12\perp}^{(0)}}{\sqrt{\gamma_1^{(0)}\gamma_2^{(0)}}}&=-\frac{3}{2}\left[\frac{\sin(k_0r_{12})}{(k_0r_{12})^2}+\frac{\cos(k_0r_{12})}{(k_0r_{12})^3}\right],\label{Omega12-perp-vac}\\
\frac{\delta_{12||}^{(0)}}{\sqrt{\gamma_1^{(0)}\gamma_2^{(0)}}}&=\frac{3}{4}\left[-\frac{\cos(k_0r_{12})}{k_0r_{12}} +\frac{\sin(k_0r_{12})}{(k_0r_{12})^2} + \frac{\cos(k_0r_{12})}{(k_0r_{12})^3}\right].\label{Omega12-para-vac}
\end{align}

Applying the energy conservation from Eqs.~(\ref{Qext})--(\ref{Qabs}) to the radiative cooperative decay rates, given in Eqs.~(\ref{Gamma12-perp-rad}) and (\ref{Gamma12-para-rad}), we finally obtain the total cooperative decay rates

\begin{align}
&\frac{\gamma_{12\perp}(k_0r_1,k_0r_2,\cos\alpha)}{\sqrt{\gamma_1^{(0)}\gamma_2^{(0)}}}=\frac{\gamma_{12\perp}^{(0)}}{\sqrt{\gamma_1^{(0)}\gamma_2^{(0)}}}\nonumber\\
&-{3}\sum_{\ell=1}^{\infty}(2\ell+1){\rm Re}\left\{a_{\ell}\left[\frac{h_{\ell}^{(1)}(k_0r_1)h_{\ell}^{(1)}(k_0r_2)}{k_0^2r_1r_2}\right]\right\}\pi_{\ell}(\cos\alpha),\label{Gamma12-perp}\\
&\frac{\gamma_{12||}(k_0r_1,k_0r_2,\cos\alpha)}{\sqrt{\gamma_1^{(0)}\gamma_2^{(0)}}}=\frac{\gamma_{12||}^{(0)}}{\sqrt{\gamma_1^{(0)}\gamma_2^{(0)}}}\nonumber\\
&-\frac{3}{2}\sum_{\ell=1}^{\infty}\frac{(2\ell+1)}{\ell(\ell+1)}{\rm Re}\Bigg\{a_{\ell}\left[\frac{\xi_{\ell}'(k_0r_1)\xi_{\ell}'(k_0r_2)}{k_0^2r_1r_2}\right]\pi_{\ell}(\cos\alpha) \nonumber\\
&+ b_{\ell} \left[h_{\ell}^{(1)}(k_0r_1)h_{\ell}^{(1)}(k_0r_2)\right]\tau_{\ell}(\cos\alpha)\Bigg\},\label{Gamma12-para}
\end{align}
where $\chi_{\ell}(z)=-z y_{\ell}(z)$ is the Riccati-Neumann function.
Hence the corresponding frequency shifts due to the dipole-dipole interaction and the sphere, for the two basic orientations, are
\begin{align}
&\frac{\delta_{12\perp}(k_0r_1,k_0r_2,\cos\alpha)}{\sqrt{\gamma_1^{(0)}\gamma_2^{(0)}}}=\frac{\delta_{12\perp}^{(0)}}{\sqrt{\gamma_1^{(0)}\gamma_2^{(0)}}}\nonumber\\
&-\frac{3}{2}\sum_{\ell=1}^{\infty}(2\ell+1){\rm Im}\left\{a_{\ell}\left[\frac{h_{\ell}^{(1)}(k_0r_1)h_{\ell}^{(1)}(k_0r_2)}{k_0^2r_1r_2}\right]\right\}\pi_{\ell}(\cos\alpha),\label{Omega12-perp}\\
&\frac{\delta_{12||}(k_0r_1,k_0r_2,\cos\alpha)}{\sqrt{\gamma_1^{(0)}\gamma_2^{(0)}}}=\frac{\delta_{12||}^{(0)}}{\sqrt{\gamma_1^{(0)}\gamma_2^{(0)}}}\nonumber\\
&-\frac{3}{4}\sum_{\ell=1}^{\infty}\frac{(2\ell+1)}{\ell(\ell+1)}{\rm Im}\Bigg\{a_{\ell}\left[\frac{\xi_{\ell}'(k_0r_1)\xi_{\ell}'(k_0r_2)}{k_0^2r_1r_2}\right]\pi_{\ell}(\cos\alpha) \nonumber\\
&+ b_{\ell} \left[h_{\ell}^{(1)}(k_0r_1)h_{\ell}^{(1)}(k_0r_2)\right]\tau_{\ell}(\cos\alpha)\Bigg\}.\label{Gamma12-para}
\end{align}

Now we are in conditions to calculate the total decay rates and the corresponding nonradiative decay rates.
For two atomic dipoles $q=\{1,2\}$, we finally have the non-radiative contributions
\begin{align}
\gamma_{q}^{\rm nrad}(k_qr_q)&=\gamma_q(k_qr_q)-\gamma_q^{\rm rad}(k_qr_q),\\
\gamma_{12}^{\rm nrad}(k_0r_1,k_0r_2)&=\gamma_{12}(k_0r_1,k_0r_2)-\gamma_{12}^{\rm rad}(k_0r_1,k_0r_2),
\end{align}
where the total decay rates are $\gamma_q=|\hat{\mathbf{d}}_q\cdot\hat{\mathbf{r}}|^2\gamma_{q\perp} + (1-|\hat{\mathbf{d}}_q\cdot\hat{\mathbf{r}}|^2)\gamma_{q||}$ and
\begin{align}
\gamma_{12}&={\rm Re}\bigg\{(\hat{\mathbf{d}}_1\cdot\hat{\mathbf{r}}_{12})(\hat{\mathbf{d}}_2^*\cdot\hat{\mathbf{r}}_{12})\gamma_{12\perp}\nonumber\\
 &+ \left[\hat{\mathbf{d}}_1\cdot\hat{\mathbf{d}}_2^*-(\hat{\mathbf{d}}_1\cdot\hat{\mathbf{r}}_{12})(\hat{\mathbf{d}}_2^*\cdot\hat{\mathbf{r}}_{12})\right]\gamma_{12||}\bigg\}.\label{Gamma12-total}
\end{align}
An analogous expression can be written for the total dipole-dipole interaction by replacing the symbol $\gamma$ with $\delta$ in Eq.~(\ref{Gamma12-total}).

\section{Intensity-intensity correlations}
\label{Second-time}

In this paper, our aim is to investigate collective properties of the emitted electromagnetic field associated with two point dipoles near a core-shell sphere.
To this end, it is convenient to calculate correlation functions that combine the single-emitter decay rate $\gamma_q$, the collective decay rate $\gamma_{12}$, and the dipole-dipole interaction $\delta_{12}$ in a single quantity.
Correlation functions such as the two-time intensity-intensity correlation function $g^{(2)}(\tau)$ allow us to study collective effects in two-atom systems as a potential source for nonclassical light fields, e.g., photon antibunching and squeezing~\cite{Ficek_PhysRep372_2002}.

The steady-state second-order correlation function is defined as
\begin{align}
g^{(2)}(\mathbf{R}_1,\mathbf{R}_2,\tau)=\lim_{t\to\infty}\frac{G^{(2)}(\mathbf{R}_1,t;\mathbf{R}_2,t+\tau)}{G^{(1)}(\mathbf{R}_1,t)G^{(1)}(\mathbf{R}_2,t+\tau)},
\end{align}
where the first-order and second-order correlation functions are, respectively,
\begin{align*}
&G^{(1)}(\mathbf{R}_q,t_q)=\langle\mathbf{E}^{-}(\mathbf{R}_q,t_q)\mathbf{E}^{+}(\mathbf{R}_q,t_q)\rangle,\\
&G^{(2)}(\mathbf{R}_1,t_1;\mathbf{R}_2,t_2)\nonumber\\
&=\langle\mathbf{E}^{-}(\mathbf{R}_1,t_1)\mathbf{E}^{-}(\mathbf{R}_2,t_2)\mathbf{E}^{+}(\mathbf{R}_2,t_2)\mathbf{E}^{+}(\mathbf{R}_1,t_1)\rangle,
\end{align*}
with $\mathbf{R}_1=R_0\mathbf{e}_{R_1}$ and $\mathbf{R}_2=R_0\mathbf{e}_{R_2}$ being the positions where the field intensities are detected at time $t_1=t$ and $t_2=t+\tau$, respectively, and $\mathbf{E}^+(\mathbf{R}_q,t_q)$ and $\mathbf{E}^-(\mathbf{R}_q,t_q)$ are the positive and negative frequency parts of the electric-field operator.
Depending on the value of $g^{(2)}(0)$, different field statistics of the emitted light can be distinguished: antibunched [$g^{(2)}(0)<1$], coherent [$g^{(2)}(0)=1$], bunched $[g^{(2)}(0)>1]$, and superbunched or extrabunched [$g^{(2)}(0)>2$]~\cite{Ficek_PhysRevA98_2018}.
Indeed, $g^{(2)}(\tau)$ is proportional to a joint probability of finding one photon around $\mathbf{R}_1$ at $t_1$ and another photon around $\mathbf{R}_2$ at $t_2$.
Hence, $g^{(2)}(0)<g^{(2)}(\tau)$ for $\tau>0$ implies photon antibunching, i.e., the joint probability of detecting two photons at the same time $t_1=t_2=t$ is smaller than at different times $t_1=t$ and $t_2=t+\tau$; conversely, for $g^{(2)}(0)>g^{(2)}(\tau)$ one has photon bunching in the emitted radiation.

In general, numerical methods are unavoidable to deal with the second-order correlation function $g^{(2)}$ for more than one atom.
Nevertheless, for the two-atom system in the stationary state, if the emitters are in equivalent positions in relation to the driving coherent field (i.e., $\mathbf{k}\cdot\mathbf{r}_{12}=0$), the steady-state atomic correlation functions can be simplified to obtain analytical solutions~\cite{Ficek_PhysRep372_2002}.
In the limiting case of weak external fields at resonance, one has a simple solution for the one-time $(\tau=0)$ second-order correlation function~\cite{Wiegand,Ficek_PhysRep372_2002}:
\begin{align}
g^{(2)}&(\mathbf{e}_{R_1},\mathbf{e}_{R_2},\tau=0)\nonumber\\
&=\frac{1}{2}\left[\left(1+\frac{\gamma_{12}}{\gamma}\right)^2+\left(\frac{2\delta_{12}}{\gamma}\right)^2\right]\nonumber\\
&\times\frac{1+\cos\left[k_0\mathbf{r}_{12}\cdot(\mathbf{e}_{R_1}-\mathbf{e}_{R_2})\right]}{\left[1+\cos(k_0\mathbf{r}_{12}\cdot\mathbf{e}_{R_1})\right]\left[1+\cos(k_0\mathbf{r}_{12}\cdot\mathbf{e}_{R_2})\right]},\label{g2-trivial}
\end{align}
where $\gamma=(\gamma_1+\gamma_2)/2$ and $\mathbf{e}_{R_1}$ and $\mathbf{e}_{R_2}$ are unit vectors of $\mathbf{R}_1$ and $\mathbf{R}_2$, respectively.
  Interestingly, note that $g^{(2)}(0)\propto|2\delta_{12} -\imath(\gamma+\gamma_{12})|^2$, i.e.,  it depends directly on the single-atom decay rates  $(\gamma)$ and the response function $(2\delta_{12} -\imath\gamma_{12})$; see Eq.~(\ref{Chi}).	
Moreover, conditions for photon-bunching or antibunching in the emitted field are readily obtained by setting $\cos(k_0\mathbf{r}_{12}\cdot\mathbf{e}_{R_q})=-1$ or $\cos[k_0\mathbf{r}_{12}\cdot(\mathbf{e}_{R_1}-\mathbf{e}_{R_2})]=-1$, respectively.
As a result, the probability for the simultaneous emission of two photons into the same direction $\mathbf{e}_{R_1}=\mathbf{e}_{R_2}$ can never become zero for a two-atom system in free space.
A similar conclusion can be obtained for the case $k_0r_{12}\ll1$, which implies $\gamma_{12}\approx\gamma$ and $\delta_{12}\gg\gamma$.
Note that for very weak external fields, one may observe $g^{(2)}(0)$ arbitrarily large for $\cos(k_0\mathbf{r}_{12}\cdot\mathbf{e}_{R_q})=-1$.
This result does not necessarily indicate very strong two-photon correlations, but rather that the probability of the emission of two single photons is much smaller than that of the simultaneous emission of two photons~\cite{Ficek_PhysRevA98_2018}.

Since we are interested in the influence of a core-shell nanosphere on the properties of the emitted field, let us set $\mathbf{e}_{R_1}=\mathbf{e}_{R_2}=\mathbf{e}_z$ and $\mathbf{r}_{12}\cdot\mathbf{e}_z=0$.
The first condition corresponds to the most commonly used configuration, where one considers a single detector to collect the fluorescence emitted by the atoms.
The second condition implies that the emitters are in equivalent positions in relation to a laser field propagating in the $z$ direction.
These assumptions lead to a simplified expression for the two-time second-order correlation function, which reads~\cite{Wiegand}:
\begin{align}
&g^{(2)}(\mathbf{e}_{z},\mathbf{e}_{z},\tau)=1 - 2\cos(\delta_{12}\tau)e^{-(\gamma+\gamma_{12})\tau/2}\nonumber\\
&+\frac{1}{4}\left[\left(1+\frac{\gamma_{12}}{\gamma}\right)^2+\left(\frac{2\delta_{12}}{\gamma}\right)^2\right]
e^{-(\gamma+\gamma_{12})\tau}\nonumber\\
 &+ e^{-(\gamma+\gamma_{12})\tau}+ \left(1+\frac{\gamma_{12}}{\gamma}\right)e^{-(\gamma+\gamma_{12})\tau/2}\nonumber\\
 &\times\left[\cos(\delta_{12}\tau)+\frac{2\delta_{12}}{(\gamma+\gamma_{12})}\sin(\delta_{12}\tau)-e^{-(\gamma+\gamma_{12})\tau/2}\right].\label{g2-tau-simp}
\end{align}
\color{black}
Again, we emphasize that we are considering the weak external field limit at resonance, i.e., the Rabi frequency can be neglected in the calculations involving the relevant decay rates and frequencies.
Hence $g^{(2)}(\tau)$ does not depend on the pump intensity in this limit, and we can focus our attention to the influence of the dipole-dipole interaction strength and the collective decay rate on the quantum system.
Indeed, within these assumptions, it is clear that Eq.~(\ref{g2-tau-simp}) depends only on $\gamma$, $\gamma_{12}$, and $\delta_{12}$ for a fixed $\tau$.
\color{black}
In particular, for a single dipole emitter one has~\cite{Wiegand}
\begin{align}
g^{(2)}(\tau)=\left(1-e^{-\gamma\tau/2}\right)^2.
\end{align}

So far we have discussed expressions of $g^{(2)}$ calculated for two emitters with the same transition frequency ($\omega_1=\omega_2$) in free space driven by a coherent resonant laser field.
Here we explore the fact that $g^{(2)}$ defined above depends only on the LDOS and the response function to include the influence of a core-shell sphere on the intensity-intensity correlations.
In order to apply Eq.~(\ref{g2-tau-simp}) to the case of two emitters near a spherical body, we impose that $|\mathbf{r}_1|=|\mathbf{r}_2|$ and $\mathbf{k}\cdot\mathbf{r}_{12}=0$, and hence $\gamma_1\approx\gamma_2$.
\color{black}
This is necessary because Eq.~(\ref{g2-tau-simp}) does not include the case of nonidentical emitters ($\omega_1\not=\omega_2$ and $\gamma_1\not=\gamma_2$), which would impose a dependence not only on $(\gamma_1+\gamma_2)$ but also on $(\gamma_1-\gamma_2)$, where the splitting between the intermediate collective states would be $\sqrt{\delta_{12}^2 + \Delta^2}$ instead of $\delta_{12}$, with $\Delta=(\omega_1-\omega_2)/2$~\cite{Ficek_PhysRep372_2002,Ficek_PhysA146_1987}.
A rigorous study of $g^{(2)}(\tau)$ calculated for two emitters with $\omega_1\not=\omega_2$ and $\gamma_1\not=\gamma_2$ in the presence of a sphere, taking into account the detuning from the incident laser beam, is beyond the scope of our manuscript and will be addressed elsewhere.
\color{black}
In addition, we assume that the orientation of the electric dipole moments $\mathbf{d}_1$ and $\mathbf{d}_2$ are determined by local electric fields at $\mathbf{r}_1$ and $\mathbf{r}_2$, respectively, which are composed by the incident field and the field scattered by the sphere~\cite{Zadkov_PhysRevA85_2012}.
These classical fields are given in Appendix~\ref{Lorenz-Mie}.

\subsection{Light scattering by a gain-assisted nanosphere coated with a plasmonic shell}
\label{sec-scattering}

\begin{figure}[htbp]
\centering
\includegraphics[width=1.1\columnwidth]{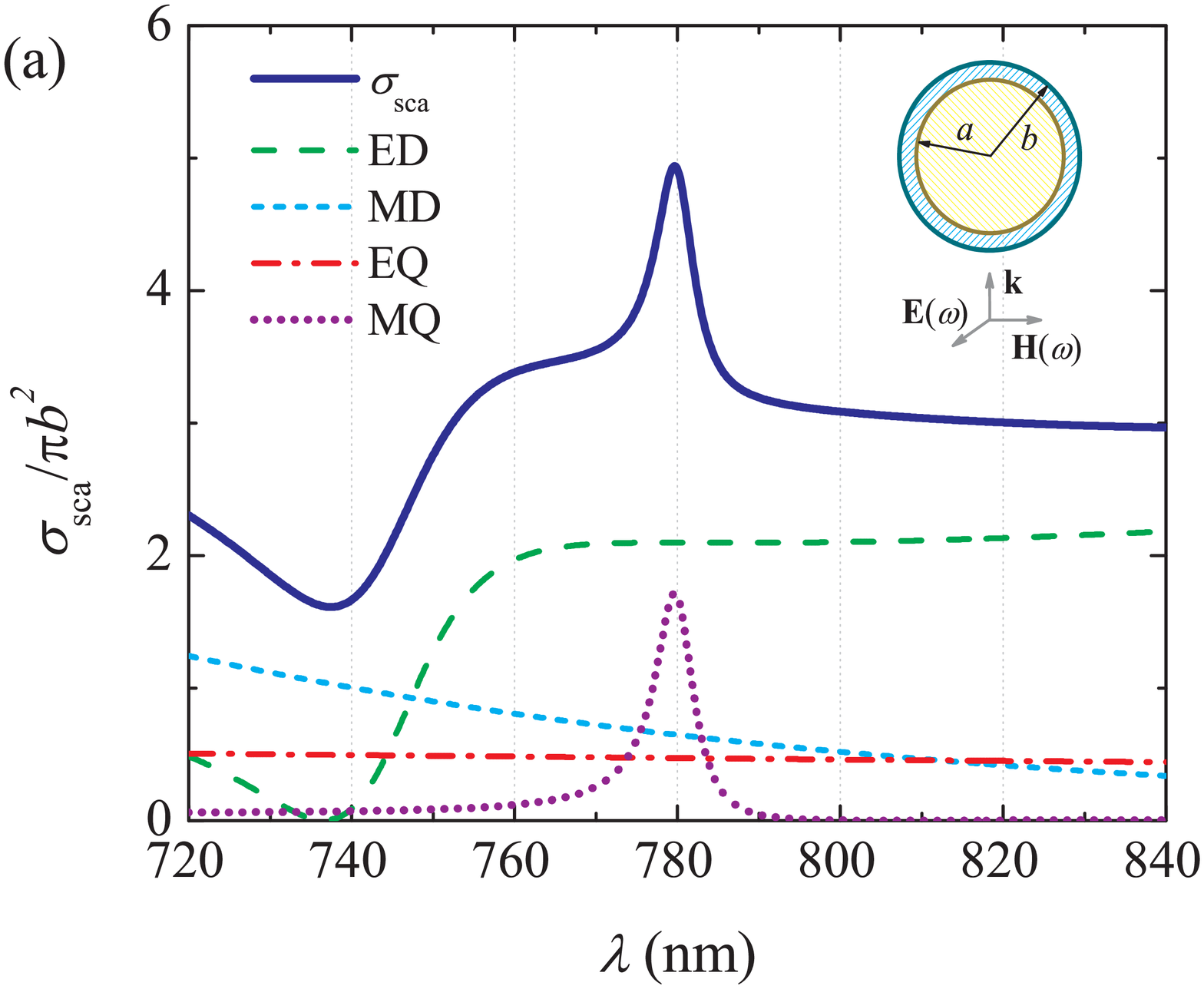}\vspace{-0.5cm}\\
\includegraphics[width=1.1\columnwidth]{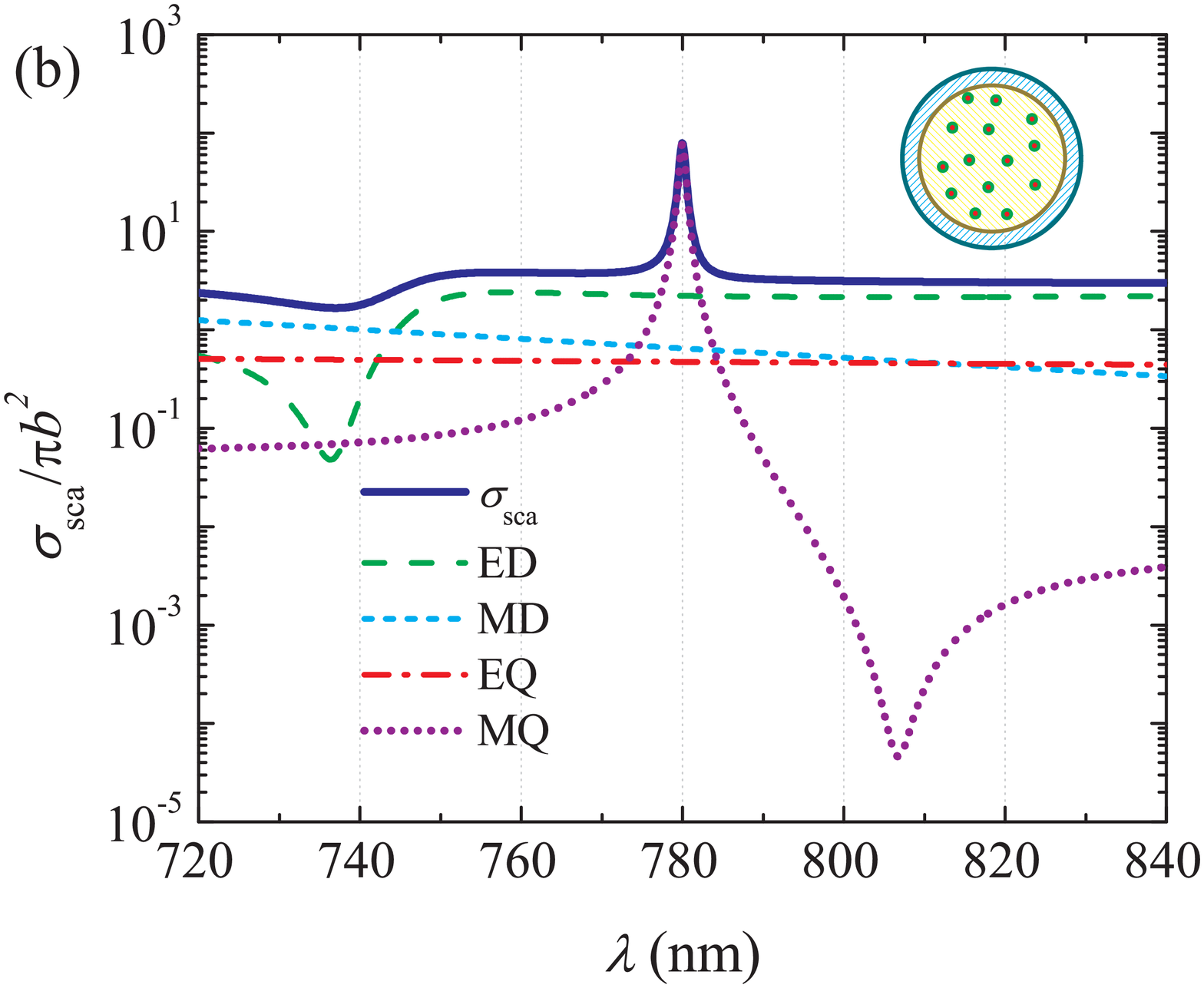}\vspace{-0.8cm}\\
\includegraphics[width=1.06\columnwidth]{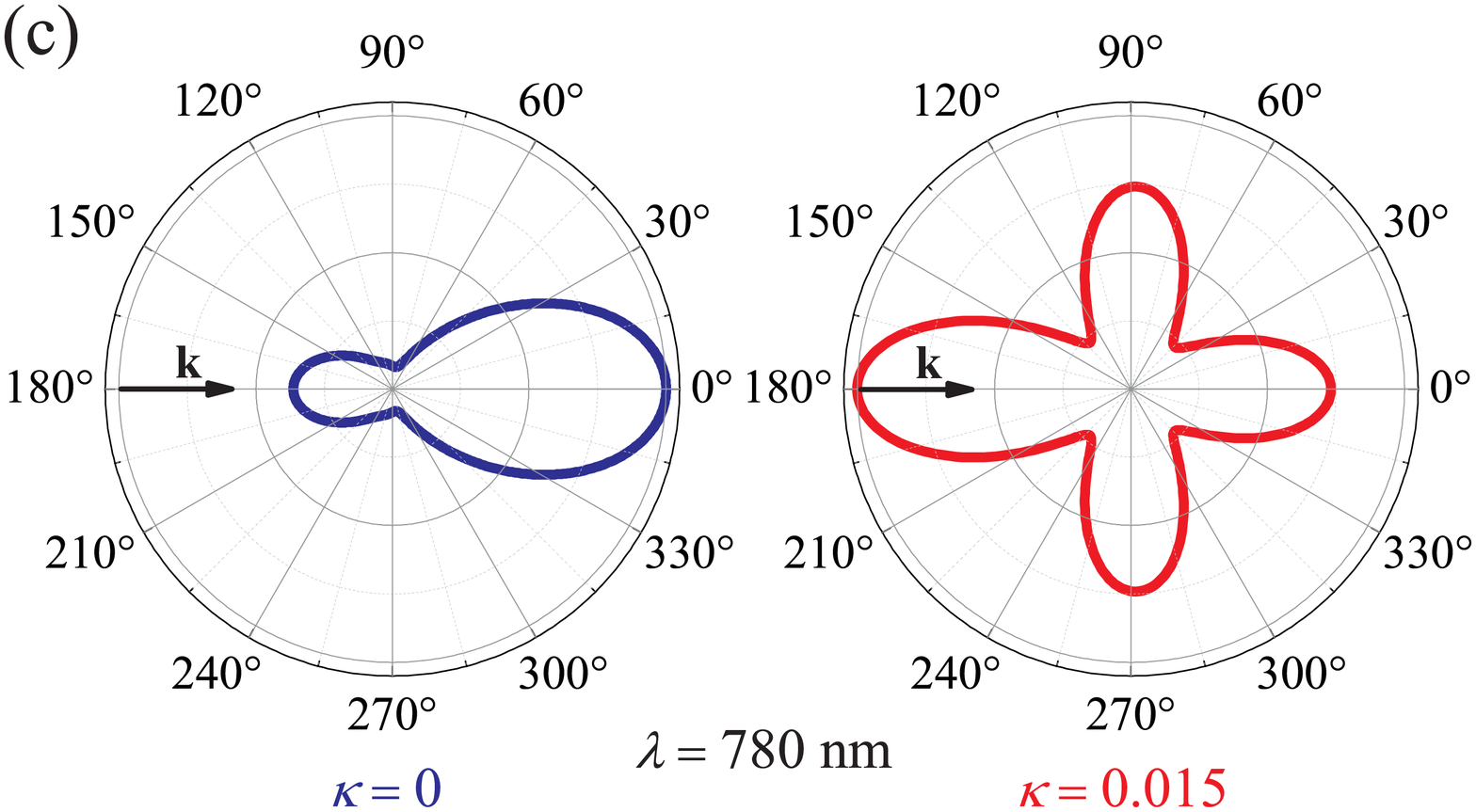}\vspace{-0.5cm}
\caption{Optical cross sections associated with light scattering by a core-shell nanosphere in free space.
A dielectric core of refractive index $n_{\rm c}=3.5 -\imath\kappa$ and radius $a=180$~nm is coated with a silver shell~\cite{Christy_PhysRevB6_1972,Shalaev_OptExp16_2008} of thickness $b-a=20$~nm.
The plots in (a) and (b) show the total scattering cross section $\sigma_{\rm sca}$  for $\kappa=0$ (no gain) and $\kappa=0.015$ (a gain-assisted core), respectively.
We show multipole contributions to the light scattering: electric dipole (ED), magnetic dipole (MD), electric quadrupole (EQ), and magnetic quadrupole (MQ).
The plot in (c) shows the differential scattering cross section associated with the MQ resonance ($\lambda=780$~nm) for $\kappa_{\rm c}=0$ (left panel) and $\kappa=0.015$ (right panel).}\label{fig2}
\end{figure}

First, let us briefly  consider the scattering properties of a silver (Ag) nanoshell coating a gain-assisted dielectric core.
Gain materials can be dielectric media doped with some dye molecules or rare-earth ions, such as Pr$^{3+}$, Ho$^{3+}$, Er$^{3+}$, Eu$^{2+}$, Nd$^{3+}$, and Tm$^{3+}$, which provide optical gain response~\cite{Tsakmakidis_Science339_2013}.
In these gain-assisted materials, there is an inversion in the number of electrons such that the population in the excited level is greater than in the lower level, which leads to a negative imaginary part of the refractive index.
In light scattering, this implies stimulated emission $(\sigma_{\rm abs}<0)$ instead of absorption $(\sigma_{\rm abs}>0)$.

Here, we consider a linear gain material consisting of doped AlGaAs, whose approximate refractive index at optical frequencies is $n_{\rm AlGaAs}=3.5-\imath\kappa$, with $\kappa$ a phenomenological gain coefficient (see Ref.~\cite{Nano} and references therein).
This linear gain approximation is valid at or below threshold~\cite{Soukoulis_PhysRevB59_1999}.
\color{black}
To describe the dielectric function of Ag, we use a Drude-Lorentz-Sommerfeld formula with an approximate interband term given by a Lorentzian tail~\cite{Shalaev_OptExp16_2008}:
\begin{equation}
\frac{\varepsilon_{\rm Ag}(\omega)}{\varepsilon_0}=1-\frac{\omega_{\rm p}^2}{\omega\left(\omega+\imath\gamma_{\infty}\right)}+\frac{f\omega_{\rm L}^2}{\omega_{\rm L}^2-\omega^2-\imath\Gamma_{\rm L}\omega},
\end{equation}
where  $\hbar\omega_{\rm p}=9.17$~eV, $\hbar\gamma_{\infty}=0.021$~eV, $\hbar\omega_{\rm L}=5.27$~eV, $\hbar\Gamma_{\rm L}=1.14$~eV, and $f=2.2$ are the parameters that provide the best fit to the Johnson and Christy experiments~\cite{Christy_PhysRevB6_1972}, for the spectral range of 380 to 1000~nm~\cite{Shalaev_OptExp16_2008}.
\color{black}
Throughout this paper, we compare two different configurations regarding the material parameters of core-shell sphere in optical frequencies: a dielectric core without gain ($\kappa=0$) and a gain-assisted dielectric core ($\kappa\not=0$).

Figure~\ref{fig2} shows the scattering cross section of a (AlGaAs) core-shell (Ag) nanosphere, with inner radius $a=180$~nm and outer radius $b=200$~nm, as a function of the wavelength.
These geometrical parameters were chosen to fulfill two conditions: (i) the scattering (plasmon) resonance occurs at $\lambda_0\approx780$~nm and (ii) the effective size of the scatter is of the order of $\lambda_0/2$.
This allows us to study the enhanced spontaneous emission rate of two single emitters with transition wavelength $\lambda_0=780$~nm in the vicinity of the sphere and its corresponding influence on the collective parameters for $r_{12}\approx\lambda_0$.

In Fig.~\ref{fig2} we show plots for the case of an inner dielectric sphere with refractive index $n_{\rm c}=n_{\rm AlGaAs}-\imath\kappa$, with ($\kappa=0.015$) and without ($\kappa=0$) gain, coated with a silver nanoshell within the Lorenz-Mie theory.
The different contributions of electric and magnetic multipoles to the total scattering cross section $\sigma_{\rm sca}$ are highlighted in Fig.~\ref{fig2}(a) and Fig.~\ref{fig2}(b).
At the resonance, note that the main contributions to the scattering are given by a broad electric dipole (ED) resonance and a narrow magnetic quadrupole (MQ) resonance at $\lambda_0\approx780$~nm, with a dominant ED response in Fig.~\ref{fig2}(a) without gain.
Due to this hierarchy of different nonvanishing multipole contributions, the overall scattering is anisotropic (Mie scattering) and the differential scattering cross section is peaked along the forward direction, remaining a dipole radiation pattern perturbed by the presence of a MQ resonance [see the left panel of Fig.~\ref{fig2}(c)].
However, when a gain medium is properly introduced within the core material ($\kappa=0.015$), it induces a stimulated emission within the core at $\lambda_0\approx780$~nm, enhancing the near-field MQ response by at least two orders of magnitude with respect to the ED response; see Fig.~\ref{fig2}(b).
Physically, this effect is due to the near-field oscillation of  plasmons excited in the inner shell interface ($r=a$) by the gain medium and the outer shell interface ($r=b$) by the incident radiation, leading to a quadrupole radiation pattern, as can be seen in the right panel of Fig.~\ref{fig2}(c).
In particular, note the presence of a Fano lineshape in the quadrupole contribution, which is a typical characteristic of interference between spectrally narrow and broad resonances in plasmonic systems~\cite{Arruda_Springer219_2018}.
Therefore, the inclusion of a gain medium inside the dielectric core can strongly amplify the plasmonic Fano resonance in the near field, leading to an enhanced ultranarrow scattering resonance in the far field.

\subsection{Gain-assisted collective spontaneous emission}
\label{Gain-assisted}

\begin{figure}[htbp!]
\centerline{\includegraphics[width=\columnwidth]{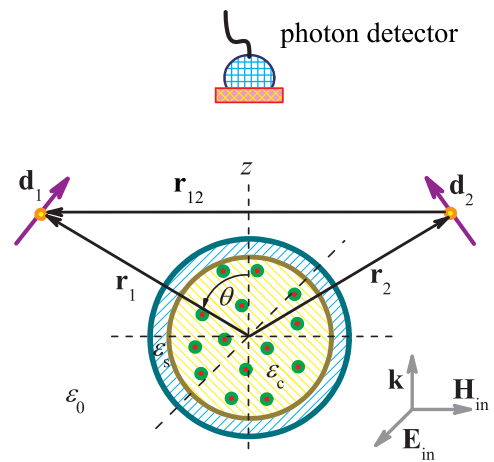}}
\caption{Two identical dipole emitters with $|\mathbf{r}_1|=|\mathbf{r}_2|$ in the vicinity of a core-shell sphere with a gain material inside.
The incoming electromagnetic wave is such that $\mathbf{k}\cdot\mathbf{r}_{12}=0$.
We consider two basic configurations: the interatomic distance vector $\mathbf{r}_{12}$ is orthogonal or parallel to the incident electric field $\mathbf{E}_{\rm in}$.
The detector is fixed in the $z$ direction.
}\label{fig3}
\end{figure}

\begin{figure*}[htbp]
\hspace{-.9cm}
\includegraphics[width=.5\textwidth]{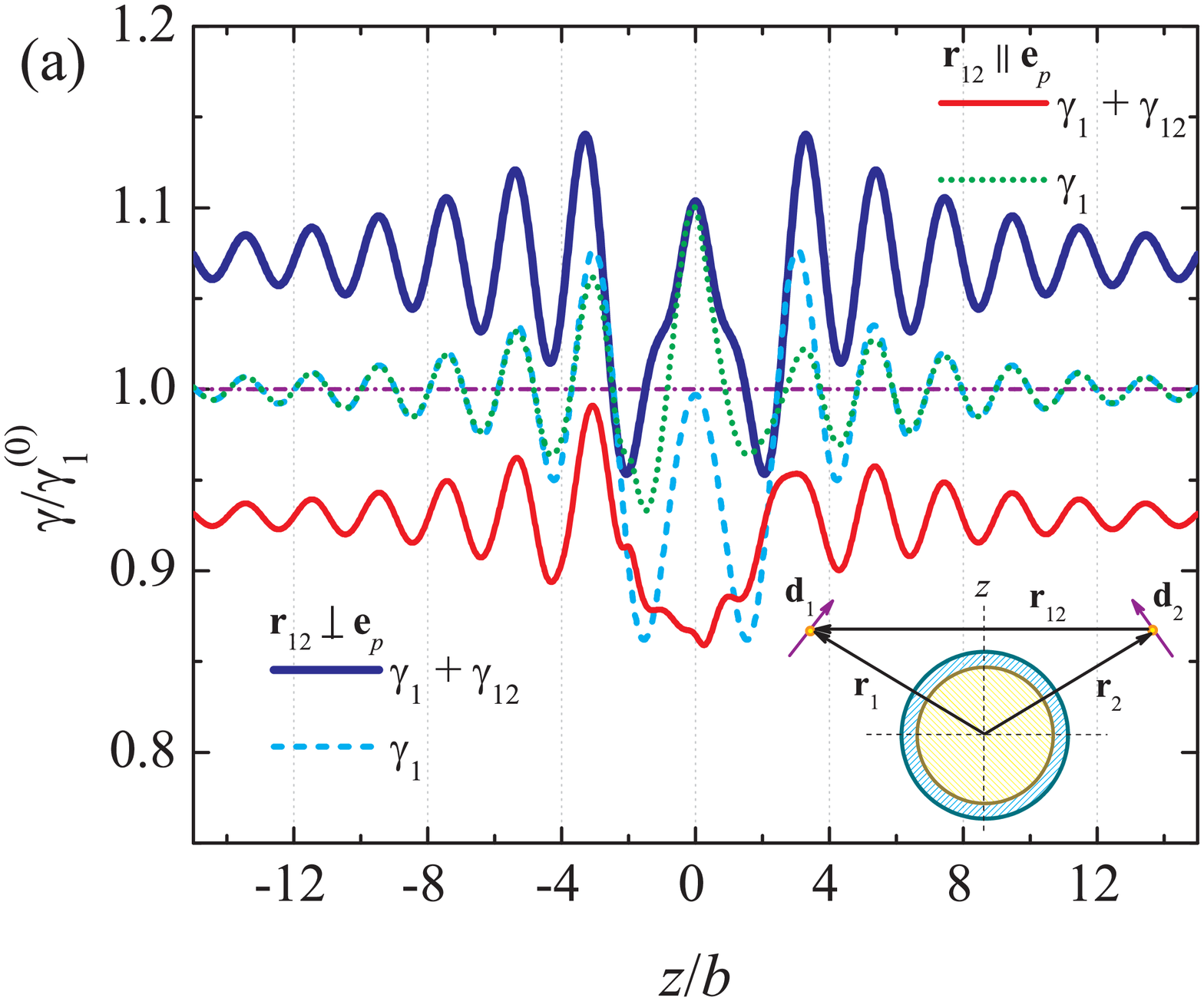}\hspace{-1cm}
\includegraphics[width=.5\textwidth]{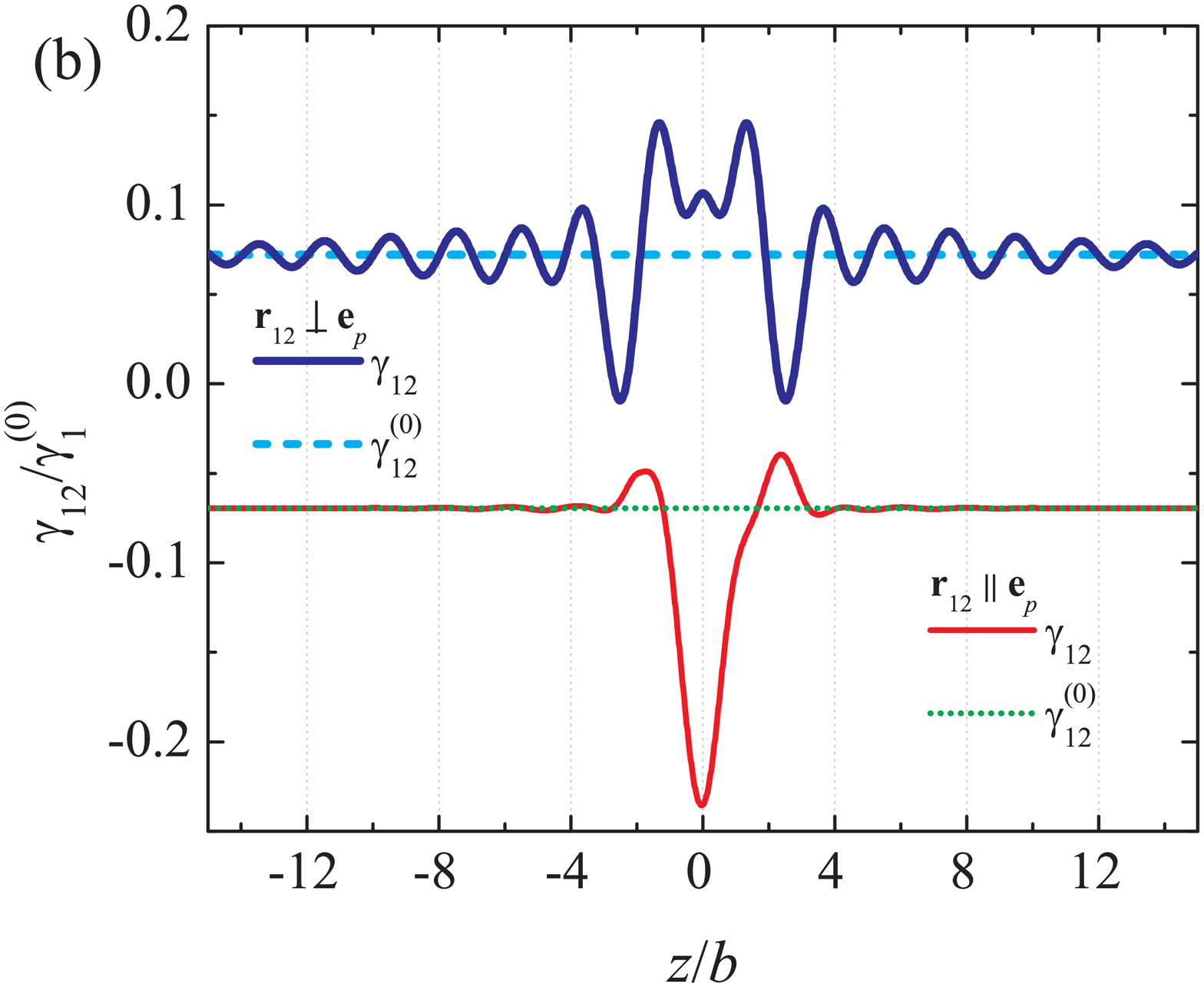}\\
\vspace{-.5cm}
\hspace{-.9cm}
\includegraphics[width=.5\textwidth]{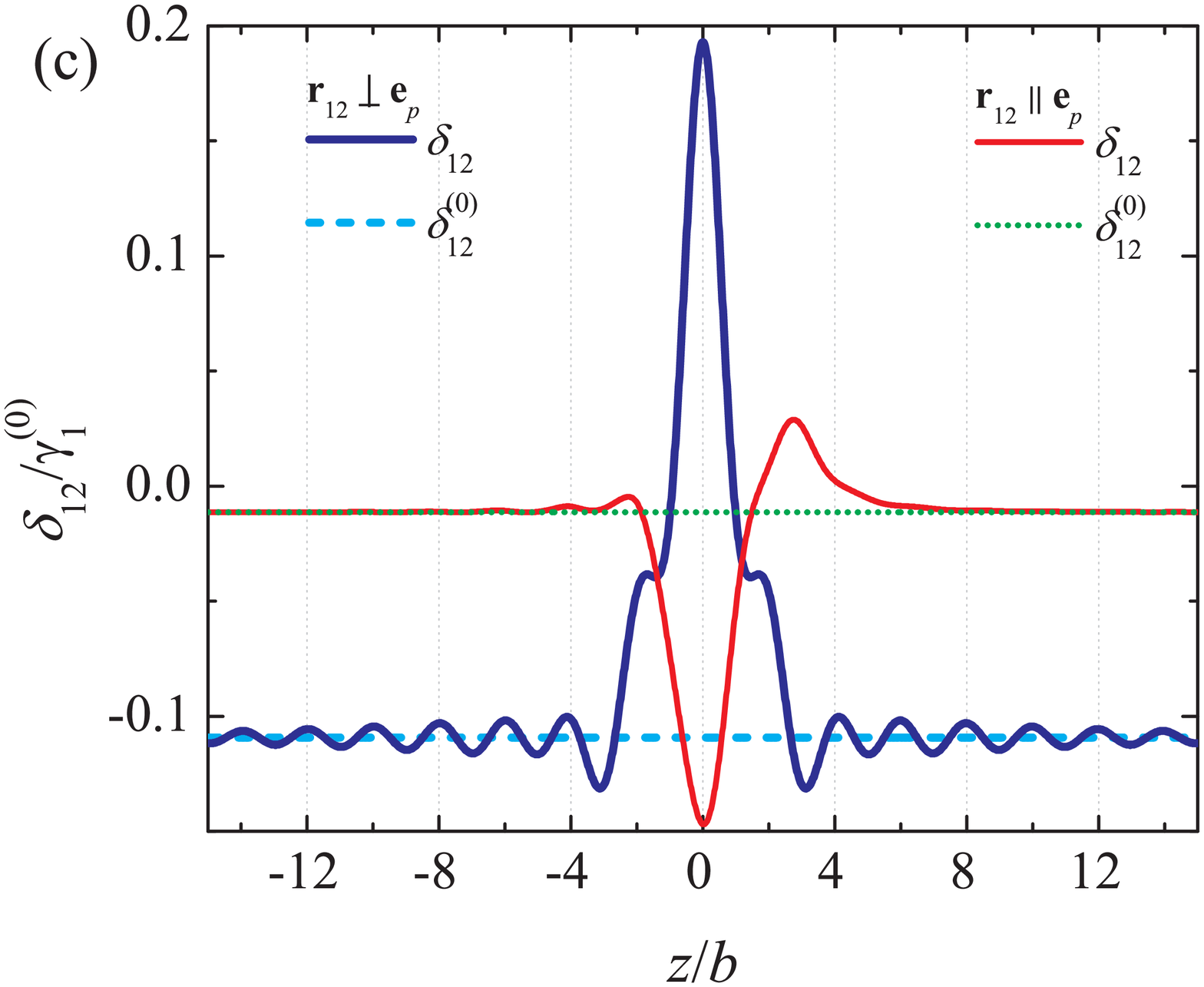}\hspace{-1cm}
\includegraphics[width=.5\textwidth]{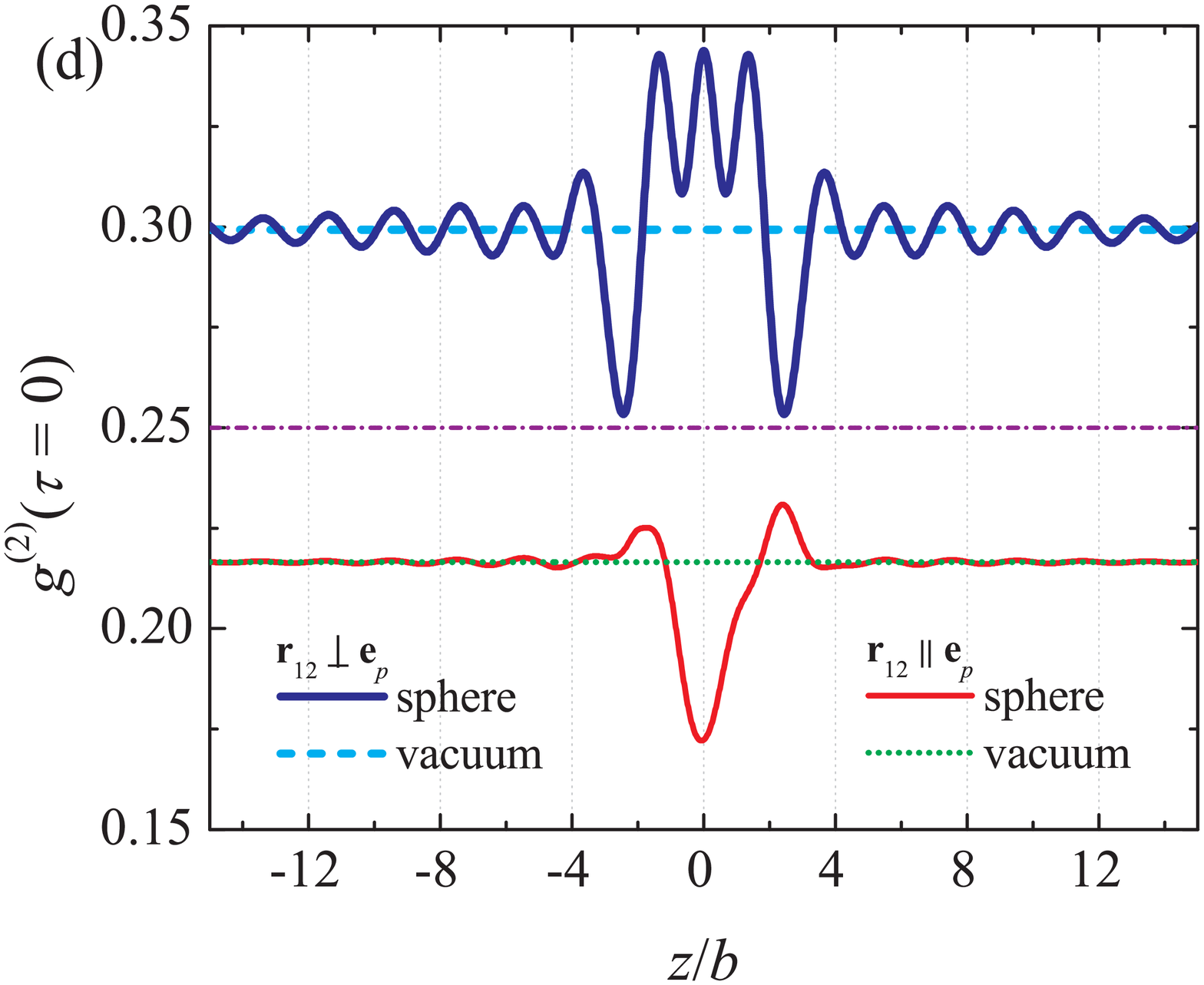}
\caption{Two dipole emitters with transition wavelength $\lambda_0=780$~nm in the vicinity of a dielectric nanosphere ($n_{\rm c}=3.5$) of radius $a=180$~nm coated with an Ag nanoshell of radius $b=200$~nm for a fixed interatomic distance $r_{12}=800$~nm.
The emitters are equally distant from the center of the sphere, where $(0,0,z)$ is the position of the midpoint of the line connecting the emitters.
$\gamma_1^{(0)}=\gamma_2^{(0)}$ is the single-emitter decay rate in free space.
The polarization of the incident electric field  $\mathbf{E}_{\rm in}$ is along $x$ direction.
(a)  Comparison between the single-emitter decay rate $\gamma_1=\gamma_2$ (with and without the cross-damping decay rate $\gamma_{12}$ contribution) in the vicinity of a coated sphere for two polarizations: $\mathbf{r}_{12}$ is parallel $(||)$ or orthogonal ($\perp$) to $\mathbf{E}_{\rm in}$.
(b) The cross-damping decay rate $\gamma_{12}$.
(c) The dipole-dipole interaction $\delta_{12}$.
(d) The normalized intensity-intensity correlation function $g^{(2)}(\tau=0)$ for detectors in $z$ direction.
The dash-dotted line corresponds to $g^{(2)}(0)$ for independent emitters ($\gamma_{12}=0=\delta_{12}$) in vacuum.}\label{fig4}
\end{figure*}

\begin{figure*}[htbp]
\hspace{-.9cm}
\includegraphics[width=.5\textwidth]{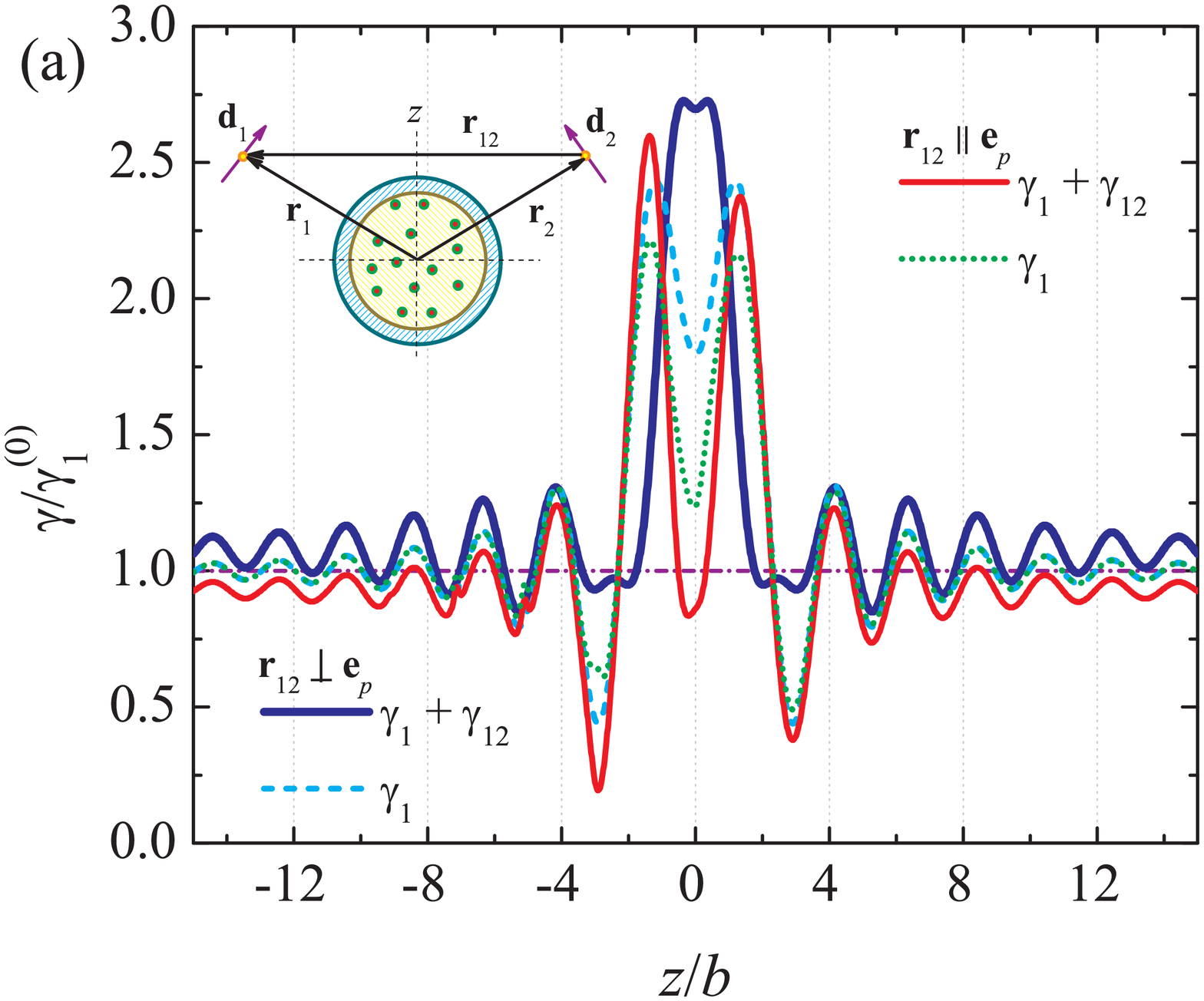}\hspace{-1cm}
\includegraphics[width=.5\textwidth]{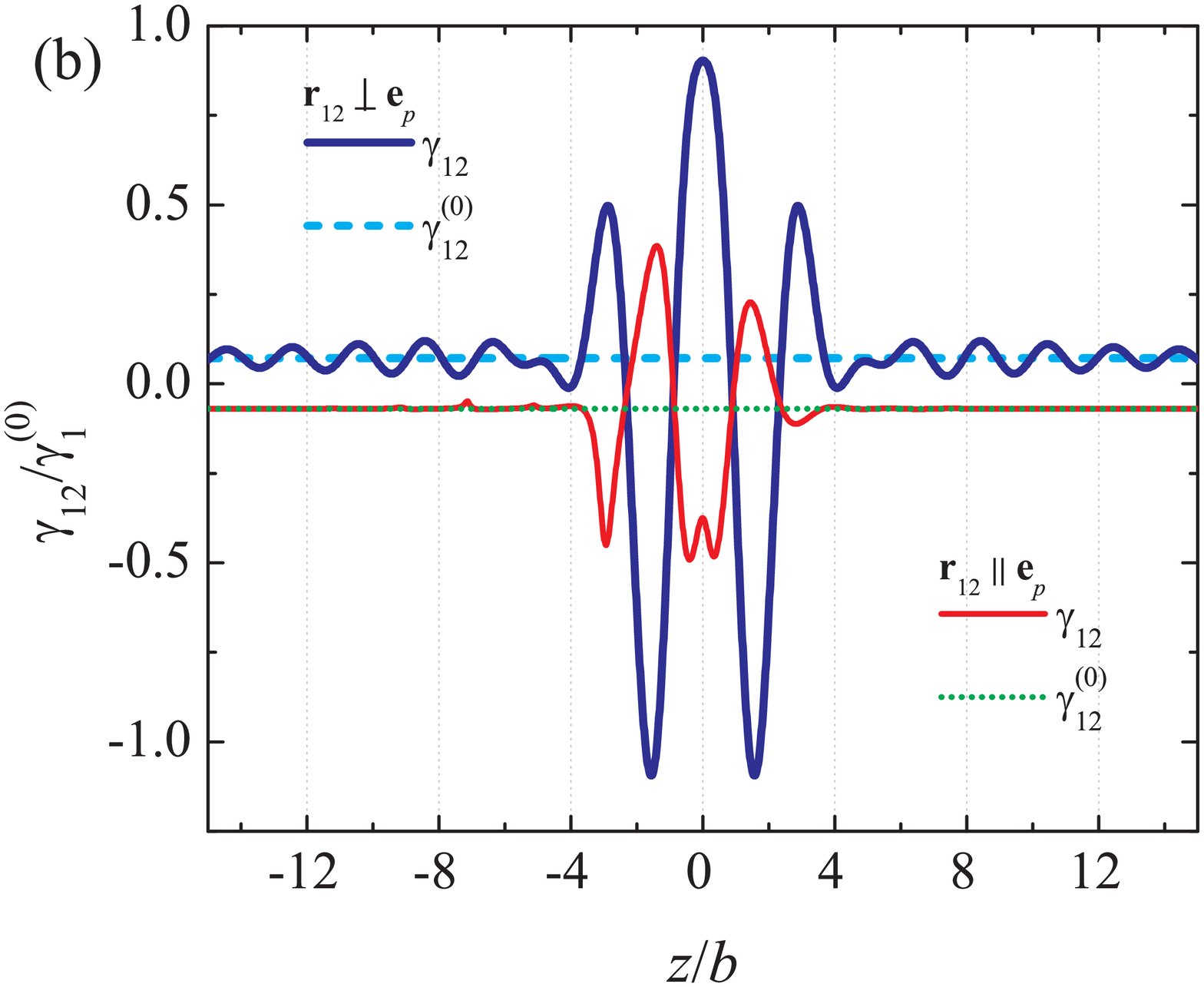}\\
\vspace{-.5cm}
\hspace{-.9cm}
\includegraphics[width=.5\textwidth]{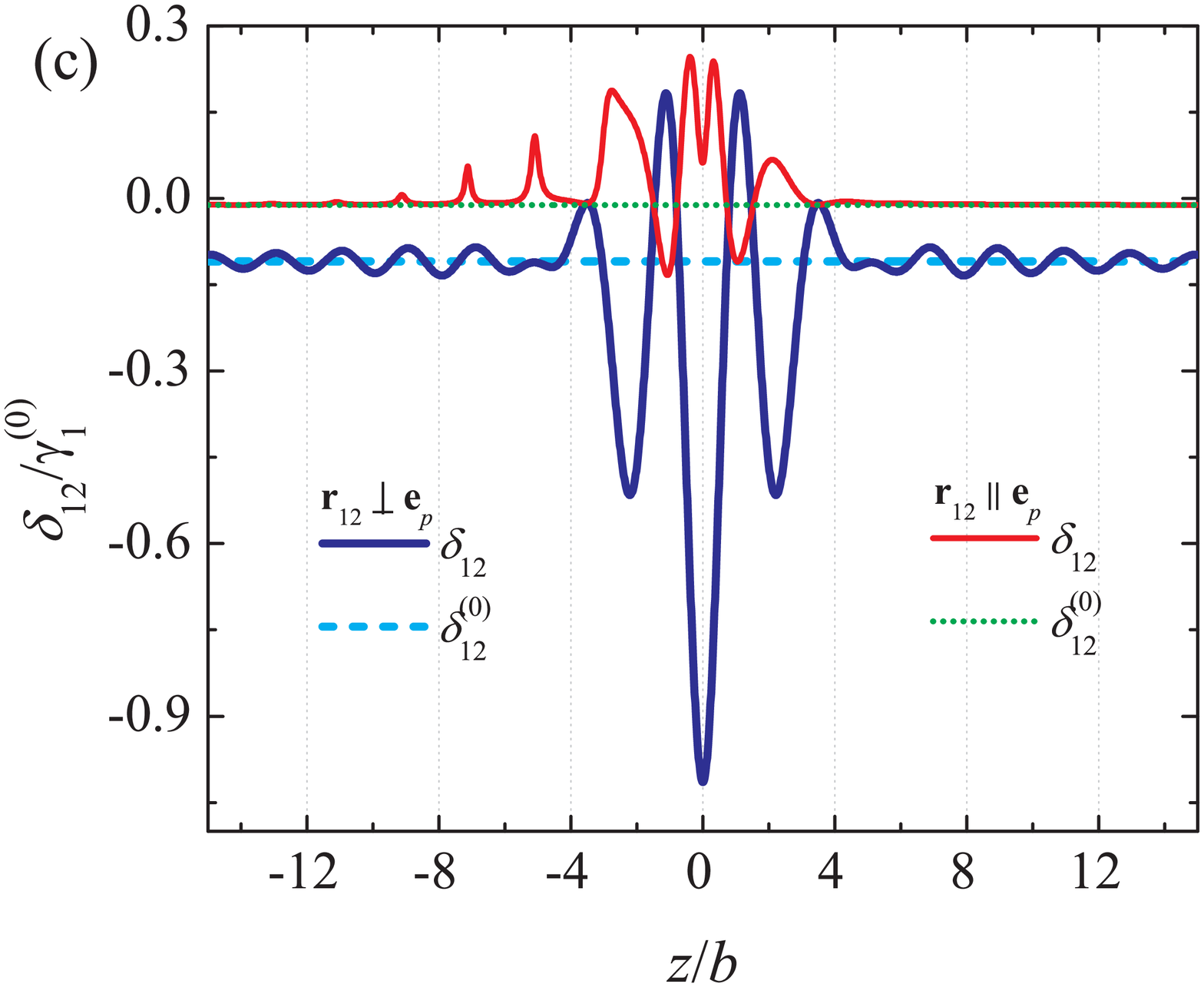}\hspace{-1cm}
\includegraphics[width=.5\textwidth]{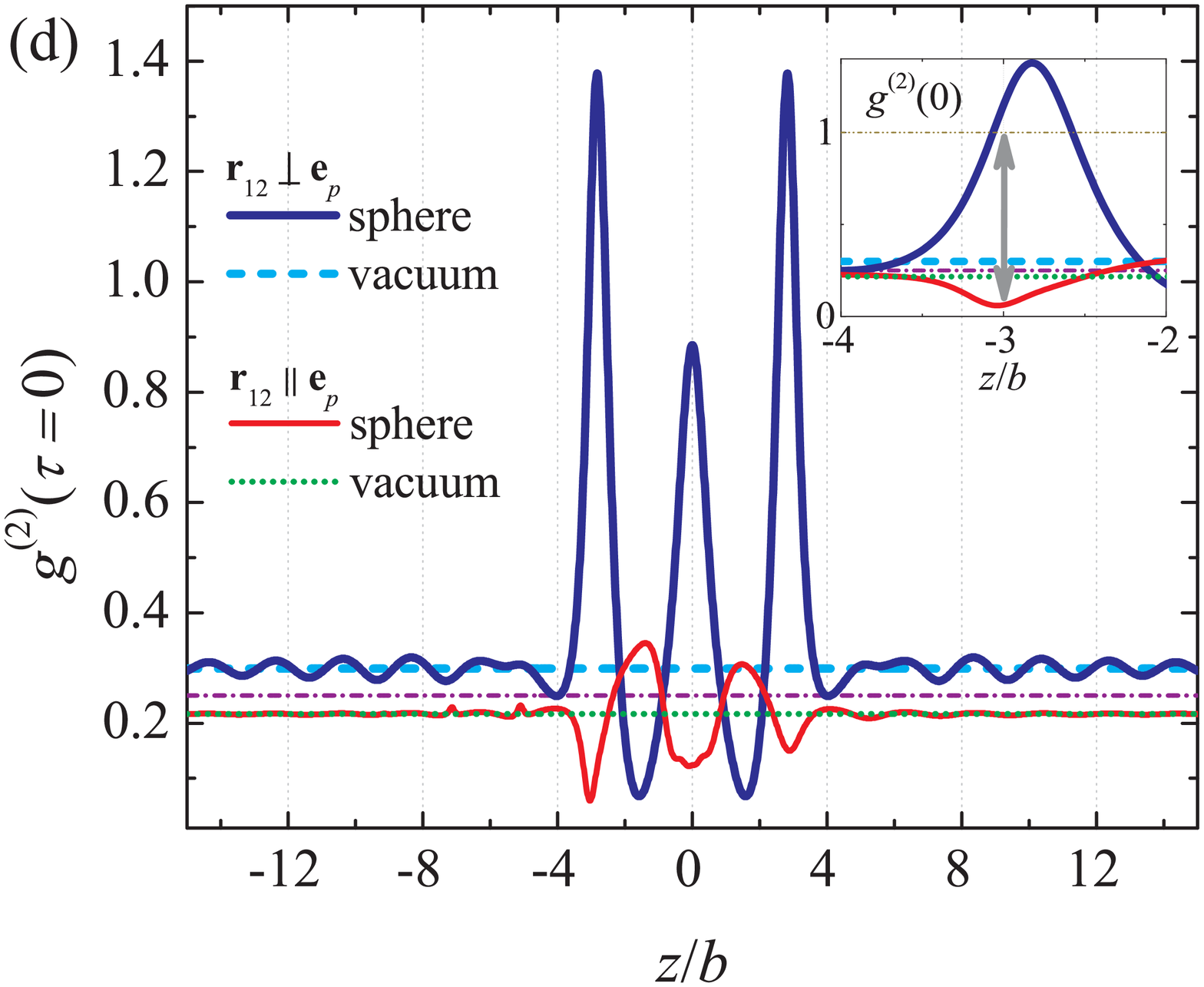}
\caption{Same quantities as in Fig.~\ref{fig4} but now with a gain-assisted core ($n_{\rm c}=3.5-\imath0.015$).
(a) Total decay rates for emitters when $\mathbf{r}_{12}$ is orthogonal ($\perp$) or parallel ($||$) to the incident electric field $\mathbf{E}_{\rm in}||{\mathbf{e}_p}$ as a function of $z$ (the position of the midpoint of the line connecting the two emitters, see inset).
(b) The cross-damping decay rate $\gamma_{12}$.
(c) The dipole-dipole interaction $\delta_{12}$.
(d) The normalized intensity-intensity correlation function $g^{(2)}(\tau=0)$ for detectors in the $z$-axis.
The inset shows the possibility of tuning $g^{(2)}(0)$ by changing the polarization of the laser beam.}\label{fig5}
\end{figure*}

\begin{figure*}[htbp]
\hspace{-.9cm}
\includegraphics[width=.5\textwidth]{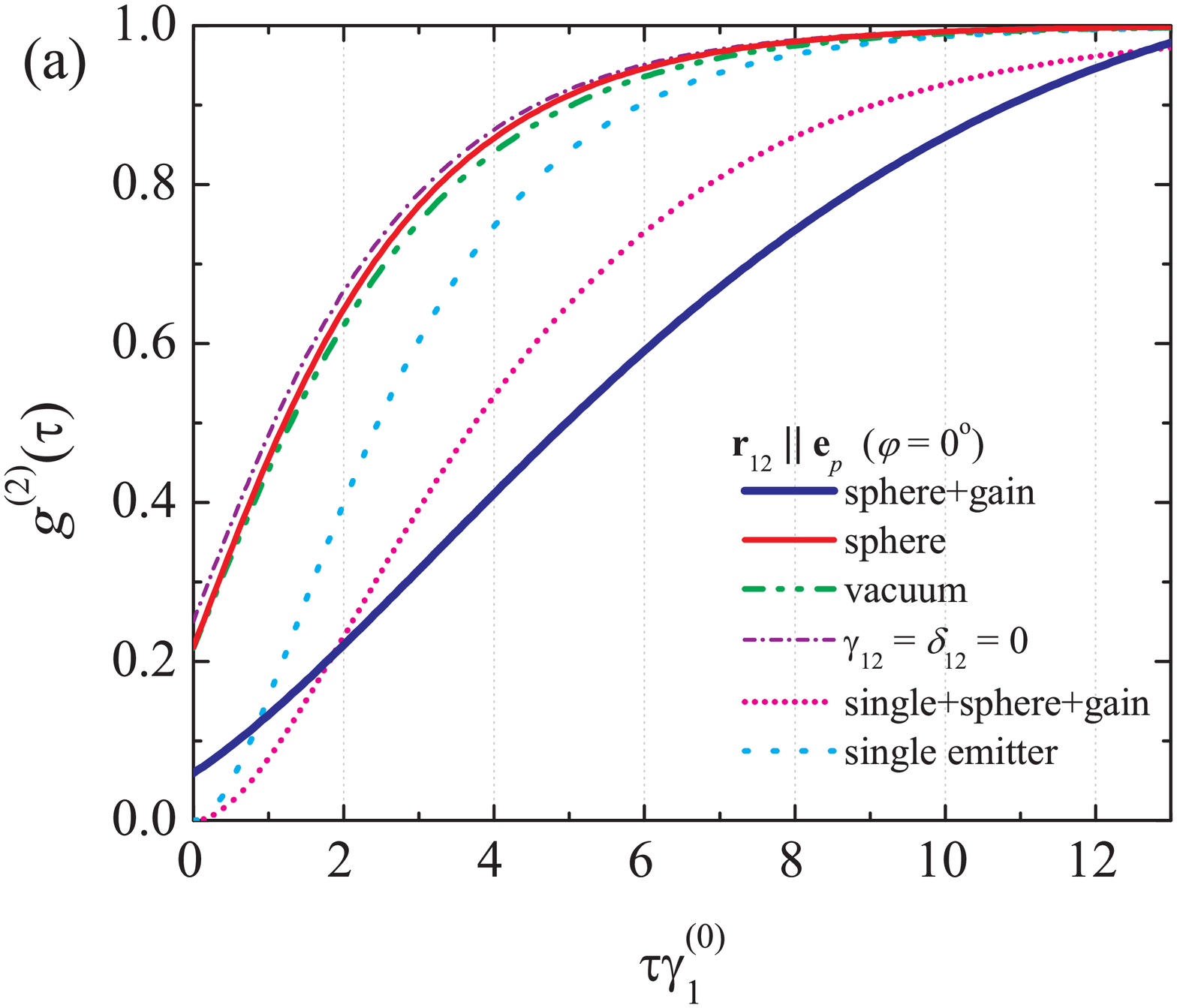}\hspace{-1cm}
\includegraphics[width=.5\textwidth]{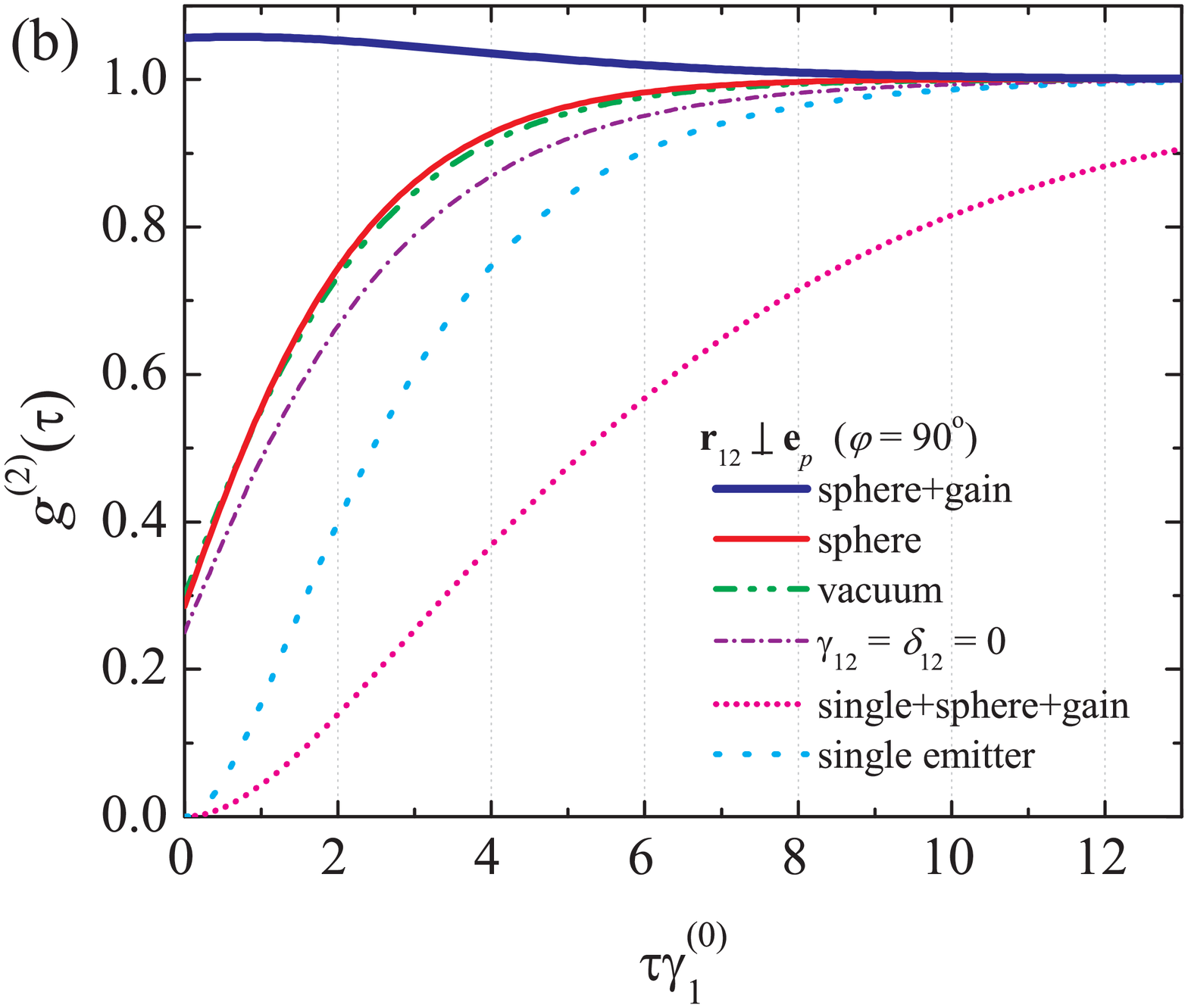}\\
\vspace{-.5cm}
\hspace{-.9cm}
\includegraphics[width=.5\textwidth]{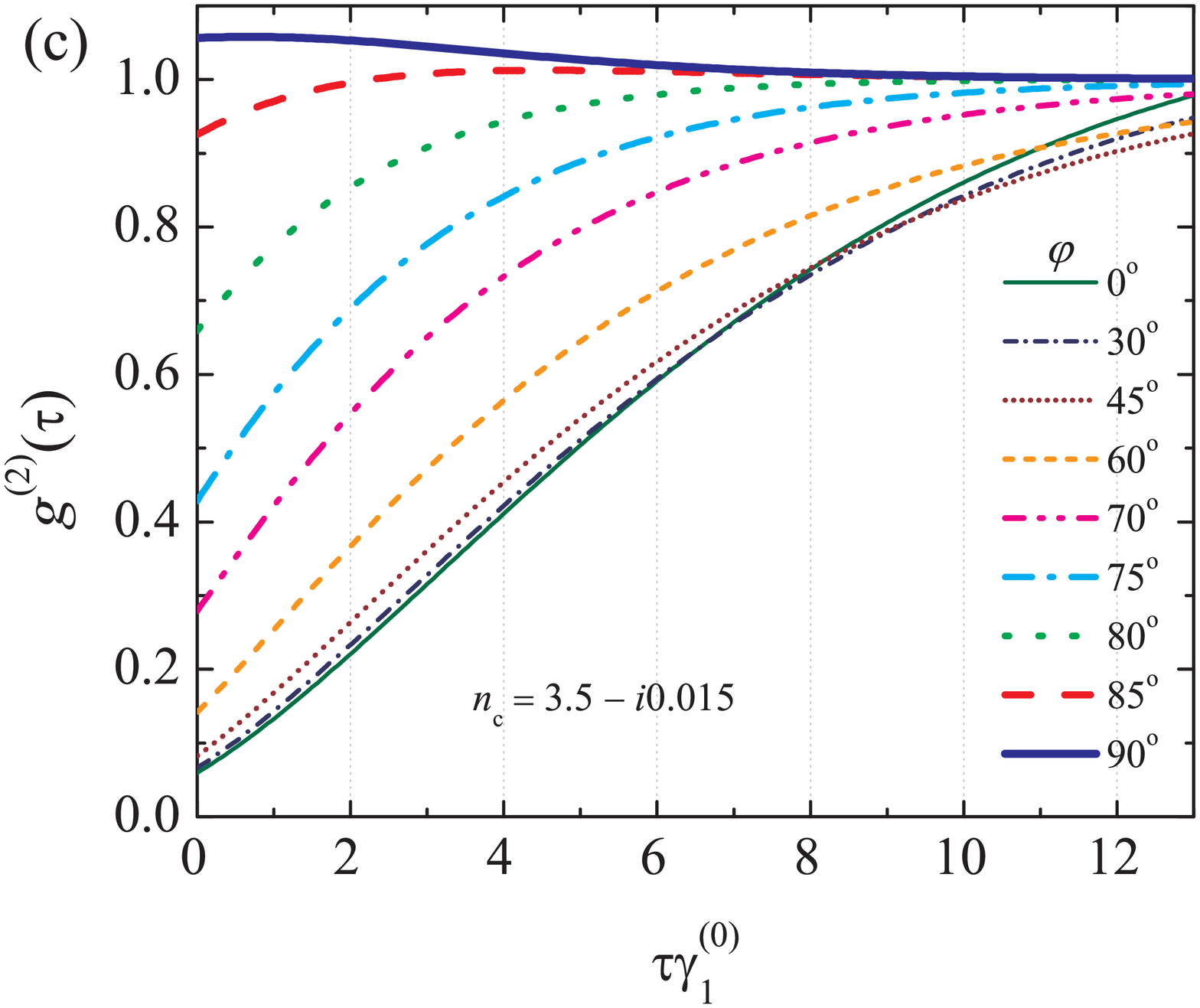}\hspace{-1cm}
\includegraphics[width=.5\textwidth]{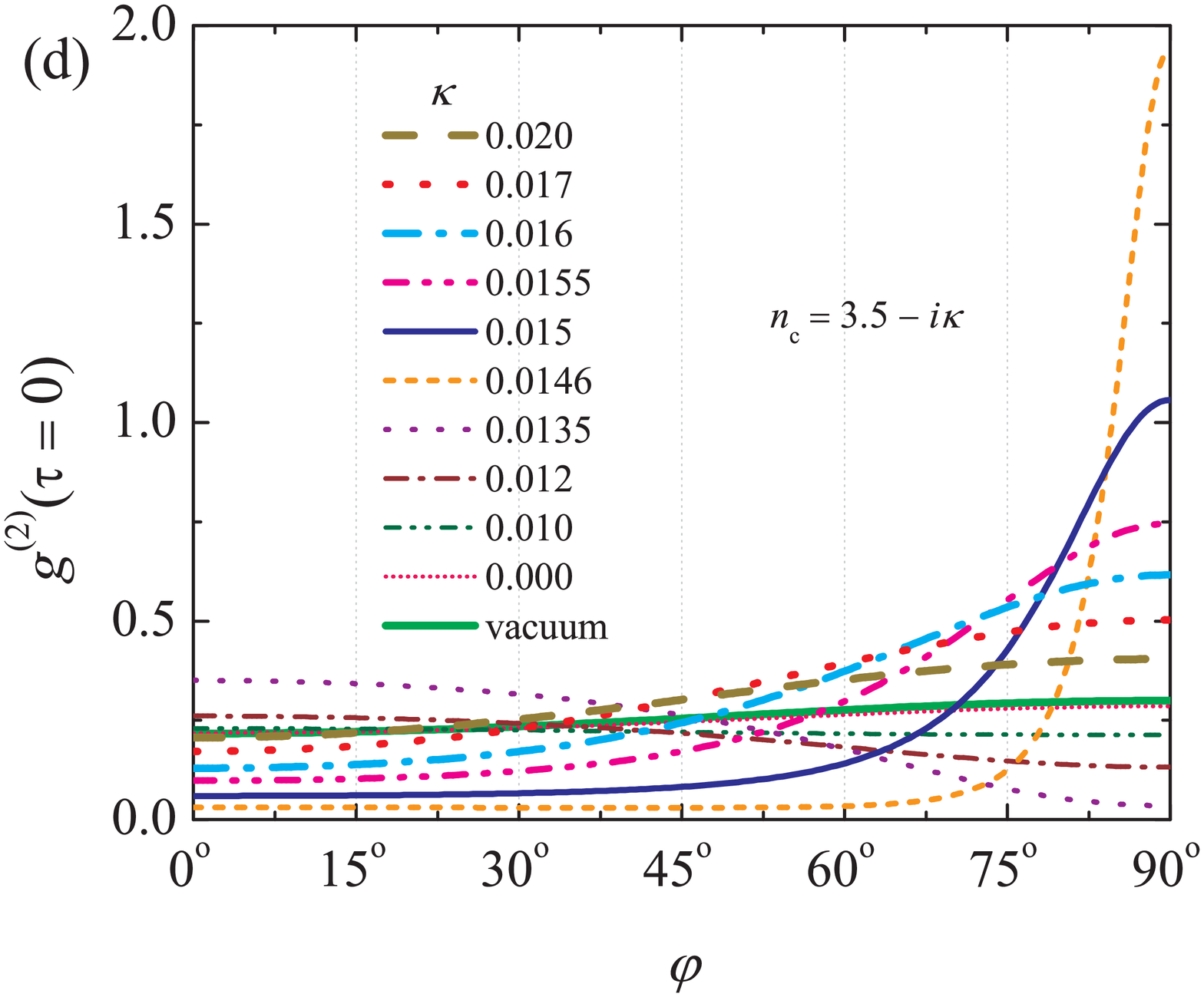}
\caption{Two-time second order correlation function $g^{(2)}(\tau)$ for two point-dipole emitters ($\lambda_0=780$~nm) near a core-shell nanosphere with the same parameters as in Fig.~\ref{fig5}: a doped AlGaAs nanosphere ($a=180$~nm) coated with an Ag layer ($b=200$~nm).
The emitters are in equivalent position in relation to the sphere and the midpoint of the interatomic distance $r_{12}=800$~nm is fixed at $z=-3.04b$.
We consider the detectors in $z$ direction.
The plots show the cases of a sphere with and without gain; two emitters in vacuum; two independent emitters in vacuum ($\gamma_{12}=\delta_{12}=0$); single emitter near a sphere with gain $(\kappa=0.015)$; and single emitter in vacuum.
$\varphi$ is the angle between $\mathbf{r}_{12}$ and $\mathbf{E}_{\rm in}||\mathbf{e}_p$.
(a) $g^{(2)}(\tau)$ for $\mathbf{r}_{12}||\mathbf{e}_p$ ($\varphi=0^{\rm o}$).
(b) $g^{(2)}(\tau)$ for $\mathbf{r}_{12}\perp\mathbf{e}_p$ ($\varphi=90^{\rm o}$).
(c) $g^{(2)}(\tau)$ for various angles $\varphi$ when the gain coefficient is $\kappa=0.015$.
(d) $g^{(2)}(0)$ as a function of $\varphi$ for different values of the gain coefficient $\kappa$.} \label{fig6}
\end{figure*}

\begin{figure}[htbp]
\includegraphics[width=1.1\columnwidth]{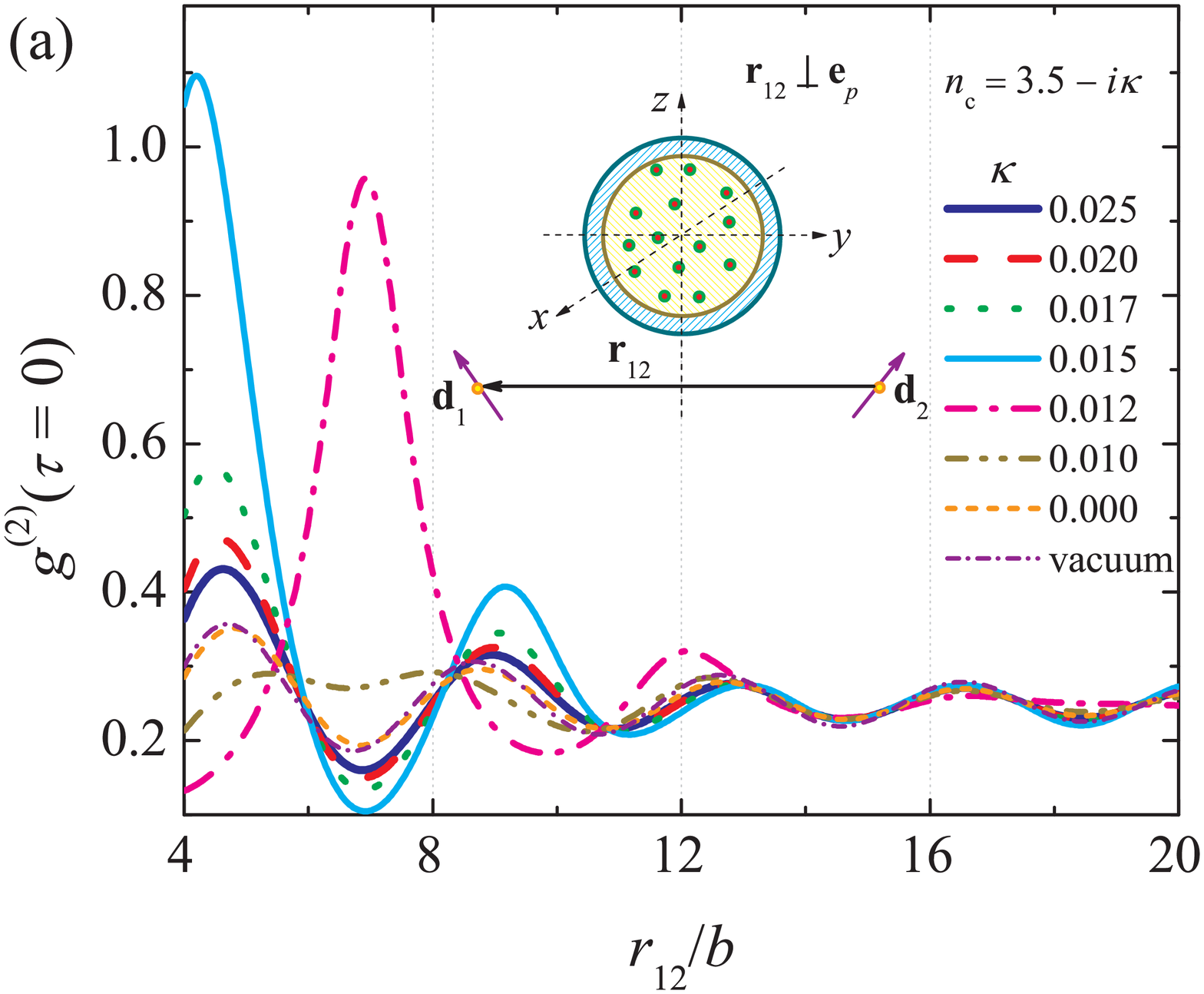}\vspace{-.5cm}
\includegraphics[width=1.1\columnwidth]{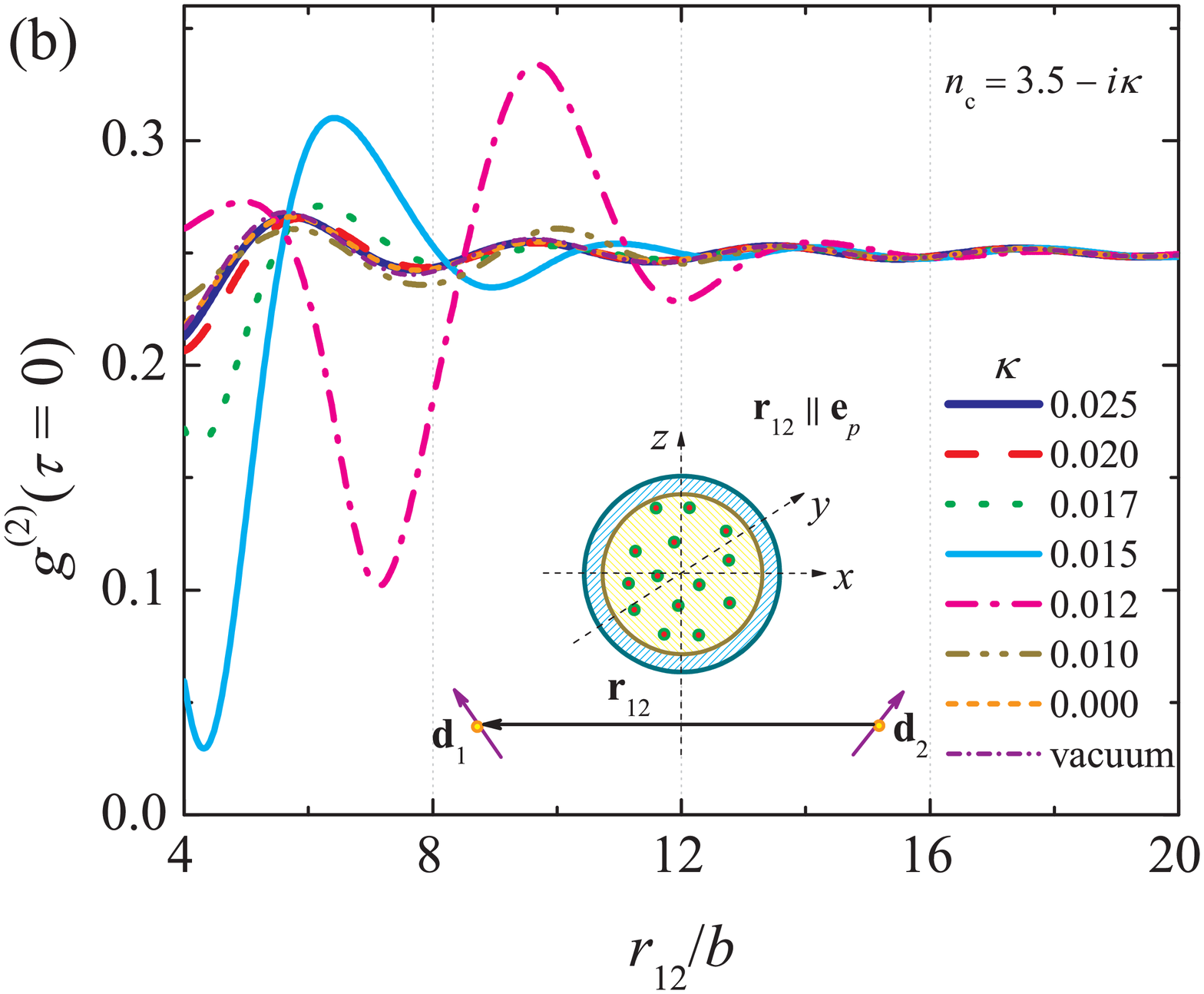}
\caption{One-time second-order correlation function $g^{(2)}(0)$ associated with two emitters in the vicinity of a gain-assisted sphere ($n_{\rm c}=3.5-\imath\kappa$) of radius $a=180$~nm coated with a silver shell of radius $b=200$~nm as a function of the interatomic distance $r_{12}$.
The midpoint of the distance $r_{12}$ is fixed at $z=-3.04b$.
The plots (a) and (b) show two different laser polarizations: $\mathbf{r}_{12}\perp\mathbf{e}_p$ and $\mathbf{r}_{12}||\mathbf{e}_p$, respectively, where $\mathbf{e}_p$ is the unity vector along the direction of the incident electric field.
The photon-bunching effect occurs for $r_{12}\approx 4b$ and $\kappa=0.015$ in (a) and can be switched to a strong antibunching by changing the polarization to (b).} \label{fig7}
\end{figure}

The change on the radiation pattern from dipole to quadrupole response in the far field in Fig.~\ref{fig2}(c) is associated with modifications to the local density of states (LDOS) in the vicinity of the core-shell sphere.
This modification to the LDOS induced by a gain material inside the core-shell sphere can be explored in the context of singe-photon sources.
The system is depicted in Fig.~\ref{fig3}.
We consider two identical dipole emitters located at positions $\mathbf{r}_1$ and $\mathbf{r}_2$ (with $|\mathbf{r}_1|=|\mathbf{r}_2|$) such that $\mathbf{k}\cdot\mathbf{r}_{12}=0$~\cite{Wiegand}.
As already discussed, these two conditions are necessary for the application of the analytical model of $g^{(2)}(\tau)$ presented in Eq.~(\ref{g2-tau-simp}).
For the sphere, the parameters are the same as before:  (AlGaAs) core-shell (Ag) nanosphere with $a=180$~nm and $b=200$~nm.
Regarding the point-dipole emitters, we consider the emission wavelength $\lambda_0=780$~nm and $r_{12}=800$~nm, for two basic configurations: $\mathbf{r}_{12}$ orthogonal or parallel to the incident electric field $\mathbf{E}_{\rm in}=E_0\mathbf{e}_p e^{\imath\mathbf{k}\cdot\mathbf{r}}$ propagating along the positive $z$ direction.
We emphasize that we only consider values of $\kappa$ for which $\gamma>0$ and $\gamma+\gamma_{12}>0$, so that Eq.~(\ref{g2-tau-simp}) can be applied.

First, consider in Fig.~\ref{fig4} the simple case of two emitters in the vicinity of a plasmonic silver nanoshell with a dielectric core ($n_{\rm c}=3.5$).
In the plots, we fix the interatomic distance $r_{12}$ and vary the position $z$ of its midpoint $(0,0,z)$.
In Fig.~\ref{fig4}(a) we see the behavior of the collective spontaneous emission rate $\gamma_1+\gamma_{12}$ (normalized to free space) when the polarization is parallel ($\mathbf{r}_{12}||\mathbf{e}_p$) or orthogonal $(\mathbf{r}_{12}\perp\mathbf{e}_p$) to the interatomic distance.
When the sphere is approximately in between the two dipole emitters ($-4b<z<4b$), we obtain different tendencies regarding the two polarizations.
When $\mathbf{r}_{12}$ is parallel to the incident electric field $\mathbf{E}_{\rm in}$, we see a small asymmetry between $z<0$ and $z>0$ and a maximum enhancement for $z=0$.
Conversely, when $\mathbf{r}_{12}$ is orthogonal to $\mathbf{E}_{\rm in}$, the Purcell factor is symmetric with respect to the sign of $z$ and presents two points of minimum Purcell factor at $z\approx\pm2b$.
Similar features appear in Figs.~\ref{fig4}(b), \ref{fig4}(c), and \ref{fig4}(d).

\color{black}
The difference in symmetry with respect to $z=0$ exhibited by the collective parameters for $\mathbf{r}_{12}||\mathbf{e}_p$ and $\mathbf{r}_{12}\perp\mathbf{e}_p$ is associated with the Mie scattering regime, i.e., the anisotropic scattering of light by the sphere.
Indeed, it is the interference of different field components with high-order multipole excitations in the vicinity of the sphere that leads to asymmetric collective parameters.
As presented in Fig.~\ref{fig2}, a sphere with radius $b$ of the same order of the wavelength $(kb\approx 1)$ exhibits high-order multipole contributions $(\ell>1)$ to the light scattering, leading to an anisotropic scattering pattern, see Fig.~\ref{fig2}~(c).
Since the electric dipole moments $\mathbf{d}_1$ and $\mathbf{d}_2$ are parallel to the local electric fields, this asymmetry between the backward ($z<0$) and forward ($z>0$) scattering directions is expected to appear in general due to a nonvanishing product of the vector projections in Eq.~(\ref{Gamma12-total}): $(\hat{\mathbf{d}}_1\cdot\hat{\mathbf{r}}_{12})(\hat{\mathbf{d}}_2^*\cdot\hat{\mathbf{r}}_{12})$.
It is precisely the product of the electric field components presented in Appendix~\ref{Lorenz-Mie} that leads to the interference of multipoles with different orders, even for emitters in equivalent positions in relation to the sphere and the incident field.
For example, considering the polarization along the $x$ direction ($\mathbf{e}_p=\mathbf{e}_x$), the spherical coordinates of the two emitters for the case $\mathbf{r}_{12}||\mathbf{e}_x$ are $\mathbf{r}_1=(r,\theta,0)$ and $\mathbf{r}_2=(r,\theta,\pi)$, with fixed $r_{12}=|\mathbf{r}_1-\mathbf{r}_2|$.
One can verify that the corresponding local electric fields have no azimuthal components for this configuration: $E_{\varphi}(\mathbf{r}_1)=E_{\varphi}(\mathbf{r}_2)=0$.
In the vicinity of the sphere, the interference of the radial and polar components of the electric field in the $xz$ plane ($\mathbf{e}_x=\sin\theta\mathbf{e}_r+\cos\theta\mathbf{e}_{\theta}$) is not invariant by changing $z$ (or $\theta$) with $-z$ (or $\pi-\theta$), which is indirectly observed in Fig.~\ref{fig4}.
Conversely, for the case $\mathbf{r}_{12}\perp\mathbf{e}_x$, one has $\mathbf{r}_1=(r,\theta,\pi/2)$ and $\mathbf{r}_2=(r,\theta,3\pi/2)$.
The radial and polar components of the local electric field vanish in this configuration: $E_{r}(\mathbf{r}_1)=E_{r}(\mathbf{r}_2)=0$ and $E_{\theta}(\mathbf{r}_1)=E_{\theta}(\mathbf{r}_2)=0$.
This means that the dipole moments only have the azimuthal component for $\mathbf{r}_{12}\perp\mathbf{e}_x$.
Since the emitters lie in the $yz$ plane ($\mathbf{e}_x=-\mathbf{e}_{\varphi}$), we have $(\hat{\mathbf{d}}_1\cdot\hat{\mathbf{r}}_{12})(\hat{\mathbf{d}}_2^*\cdot\hat{\mathbf{r}}_{12})=0$ and $\hat{\mathbf{d}}_1\cdot\hat{\mathbf{d}}_2^*=1$, and hence the collective parameters are symmetric in relation to $z=0$.
\color{black}

In Fig.~\ref{fig4}(b) and \ref{fig4}(c), we see that both the cross-damping decay rate $\gamma_{12}$ and the dipole-dipole interaction $\delta_{12}$ have different signs depending on the polarization, with strong variations when the point dipoles are very close to the sphere.
More importantly, we see in Fig.~\ref{fig4}(d) the modulation of $g^{(2)}(0)$ as a function of the polarization.
At $z=0$, i.e., when the sphere is at the midpoint between the two emitters, $g^{(2)}(0)$ is reduced by half by changing from $\mathbf{r}_{12}\perp\mathbf{e}_p$ [$g^{(2)}(0)\approx0.35$] to $\mathbf{r}_{12}||\mathbf{e}_p$ [$g^{(2)}(0)\approx0.17$].
Note that a change in $g^{(2)}(0)$ is also observed when the emitters are in vacuum.
The value of $g^{(2)}(0)$ does not depend on the polarization only for independent emitters [$g^{(2)}(0)=0.25]$.
However, the presence of the core-shell sphere increases $g^{(2)}(0)$ for $\mathbf{r}_{12}\perp\mathbf{e}_p$ and decreases $g^{(2)}(0)$ for $\mathbf{r}_{12}||\mathbf{e}_p$ when compared with the free space configuration.
In this case of $\kappa=0$, the influence of the near field in the vicinity of the sphere is not enough to change the system from $g^{(2)}(0)>1$ (photon bunching) to $g^{(2)}(0)<1$ (antibunching).

The collective parameters drastically change with the introduction of a gain medium within the dielectric core, see Fig.~\ref{fig5}.
All parameters are generally enhanced for point dipoles close to a gain-assisted nanosphere ($n_{\rm c}=3.5-\imath0.015$), presenting huge variations in the range $-4b<z<4b$.
For instance, we can see both the enhancement ($z\approx0$) and a strong suppression ($z\approx-3b$) of the spontaneous emission in Fig.~\ref{fig5}(a); change in sign of the cross-damping decay rate $\gamma_{12}$ in Fig.~\ref{fig5}(b) from $-\gamma_1^{(0)}$ to $\gamma_1^{(0)}$; and moderate values of dipole-dipole interaction in Fig.~\ref{fig5}(c).
These modifications are related to the amplification of near-field interactions induced by the gain medium (EQ and MQ excitations, see Fig.~\ref{fig2}).
By combining these gain-induced modifications into the one-time second order correlation function, we show in Fig.~\ref{fig5}(d) that we can achieve photon-bunching $[g^{(2)}(0)>1]$ and a strong photon antibunching $[g^{(2)}(0)\ll1]$ without changing the position of detectors or considering $k_0r_{12}\ll1$, which are the trivial configurations, see Eq.~(\ref{g2-trivial}).
Instead, photon-bunching or antibunching properties of the emitted field can be obtained by properly setting the distance between emitters and sphere for $r_{12}\approx\lambda_0$.

The different behaviors concerning the two polarizations of the incident laser beam show that it is possible to switch from classical to nonclassical emitted light by simply changing the polarization, where emitters, sphere and detectors are in fixed positions.
For instance, in Fig.~\ref{fig5}(d) the position $z\approx-3.04b$ shows photon bunching for point dipoles when $\mathbf{E}_{\rm in}\perp\mathbf{r}_{12}$ and a strong photon antibunching when $\mathbf{E}_{\rm in}||\mathbf{r}_{12}$.
This difference for the two polarizations is related to the quadrupole pattern of the scattering intensity exhibited by the sphere with a gain-assisted core, see Fig.~\ref{fig2}(c).

To make this point clear, in Fig.~\ref{fig6} we investigate the photon-bunching and antibunching properties in the emitted field by using the two-time second order correlation function $g^{(2)}(\tau)$ for $z=-3.04b$.
The plots highlight the differences between some of the configurations studied here.
In Fig.~\ref{fig6}(a) and Fig.~\ref{fig6}(b) we show the case of two dipole emitters in the vicinity of a sphere with and without gain (solid lines); in vacuum with and without dipole-dipole interaction (dot-dashed lines); a single emitter in the vicinity of a sphere with gain (short-dotted line); and a single emitter in vacuum (dotted line).
For the sake of clarity, we introduce the angle $\varphi$ between the interatomic distance vector $\mathbf{r}_{12}$ and the polarization vector $\mathbf{e}_p$.
As can be seen in Fig.~\ref{fig6}(a), $g^{(2)}(\tau)$
for a core-shell sphere with gain $(\kappa=0.015)$ is smaller than $g^{(2)}(\tau)$ associated with a single emitter in vacuum for $\tau>1/\gamma_1^{(0)}$ when $\varphi=0^{\rm o}$.
Conversely, we clearly see photon bunching only for the case of two emitters in the vicinity of a core-shell sphere with gain $(\kappa=0.015)$ when $\varphi=90^{\rm o}$.
Indeed, for a fixed gain coefficient $\kappa=0.015$, we show in Fig.~\ref{fig6}(b) that one could continuously change from photon bunching to antibunching by varying the angle $\varphi$.
The value of the gain must be carefully chosen according to the geometric parameters of the system, as can be verified in Fig.~\ref{fig6}(d).
For example, note that a value of $\kappa=0.0146$ provides an even higher contrast than $\kappa=0.015$ for $g^{(2)}(0)$ as a function of $\varphi$.
Here, we emphasize that we are considering values of the gain coefficient $\kappa$ that only provide positive decay rates $\gamma$ and $\gamma+\gamma_{12}$, so that Eq.~(\ref{g2-tau-simp}) is valid.

To show the dependence of $g^{(2)}(0)$ on the geometric parameters, we set $a=180$~nm, $b=200$~nm and vary the interatomic distance $r_{12}$ for various values of $\kappa$.
Once again, the emitters are in equivalent position in relation to the sphere $(r_1=r_2)$, where the midpoint of the line connecting the emitters is fixed at $z=-3.04b$.
Depending on the interatomic distance $r_{12}$, different values of $\kappa$ may induce greater variations on $g^{(2)}(0)$ than for $\kappa=0.015$ as a function of the incident polarization.
For instance, for $r_{12}\approx 6.9b$ and $\kappa=0.012$, we obtain $g^{(2)}(0)\approx0.96$ for $\mathbf{r}_{12}\perp\mathbf{E}_{\rm in}$ in Fig.~\ref{fig7}(a) and  $g^{(2)}(0)\approx0.10$ for $\mathbf{r}_{12}||\mathbf{E}_{\rm in}$ in Fig.~\ref{fig7}(b).
This is due to the fact that different values of $\kappa$ change both phase and direction of the electric dipole moments $\mathbf{d}_1$ and $\mathbf{d}_2$ in the vicinity of the sphere, and hence $g^{(2)}(0)$.
If we also vary $z$, other values of $\kappa$ will be suitable for switching the values of $g^{(2)}(0)$.
Hence, a comprehensive study of the geometrical parameters of a given system containing a gain material could provide an optimal configuration for switching between photon-bunching and antibunching effect.

\section{Conclusion}
\label{Conclusion}

In conclusion, we have investigated theoretically the collective spontaneous emission of two point-dipole emitters near a plasmonic core-shell nanosphere containing a linearly amplifying medium.
We have derived closed analytical expressions for both the cross-damping decay rate and the dipole-dipole interaction strength associated with two atomic dipoles with arbitrary position and orientation in relation to a sphere.
Using a simplified model for $g^{(2)}(\tau)$ valid for dipole emitters in equivalent positions in relation to an incident laser beam, we have shown the possibility of alternating between photon-bunching and antibunching effect as a function of the polarization of light and the position of emitters.
We have suggested that this can be achieved by properly introducing a gain material inside a plasmonic nanoshell placed in the vicinity of point-dipole emitters (e.g., quantum dots).
This result could be of interest to technological applications using polarization-dependent single-photon sources in linearly amplifying artificial medium.

\section*{Acknowledgments}

The authors acknowledge the bilateral project CAPES-DAAD Probral No. 488/2018, Process No. 88881.143936/2017-01.
T.J.A., R.B., and Ph.W.C. hold grants from S\~ao Paulo Research Foundation (FAPESP), Grant Nos. 2015/21194-3, 2018/15554-5, and 2013/04162-5, respectively.
S.S. acknowledges funding by the Deutsche Forschungsgemeinschaft (DFG, German Research Foundation), Grant No. 422447846.

\appendix

\section{Hamiltonian of the two atom-field system}
\label{Hamiltonian}

In presence of an external electromagnetic field, the Hamiltonian of the two atom-field system in the electric dipole approximation is
\begin{align}
\hat{\mathcal{H}}=\sum_{q=1}^{2}\left[\hat{\mathcal{H}}_{\rm atom}^{(q)} + \hat{\mathcal{H}}_{\rm int}^{(q)}\right] + \hat{\mathcal{H}}_{\rm field},
\end{align}
where $\hat{\mathcal{H}}_{\rm atom}^{(q)}=\hbar\omega_q\hat{S}_q^z$ is the atomic Hamiltonian, with $\hat{S}_q^z=(|{\rm e}_q\rangle\langle{\rm e}_q|-|{\rm g}_q\rangle\langle{\rm g}_q|)/2$ being the energy operator of the $q$th atom; $\hat{\mathcal{H}}_{\rm field}=\int_V{\rm d}^3r[\varepsilon_0\hat{\mathbf{E}}^2(\mathbf{r})+\mu_0^{-1}\hat{\mathbf{B}}^2(\mathbf{r})]/2$ is the electromagnetic field Hamiltonian; and $\hat{\mathcal{H}}_{\rm int}^{(q)}=-\hat{\mathbf{d}}_q\cdot\hat{\mathbf{E}}(\mathbf{r}_q)$ is the atom-field interaction Hamiltonian, with $\mathbf{r}_q$ being the position of the atom $q=\{1,2\}$~\cite{Agarwal_PhysRevA45_1992,Milonni_Book1994}.
The electric dipole moment operator satisfies $\langle{\rm e}_q|\hat{\mathbf{d}}_q|{\rm e}_q\rangle=\mathbf{0}=\langle{\rm g}_q|\hat{\mathbf{d}}_q|{\rm g}_q\rangle$ and has nonvanishing off-diagonal elements, i.e., the eigenstates have no permanent dipole moment.
We define the dipole-moment matrix element as $\mathbf{d}_{q}\equiv\langle{\rm g}_q|\hat{\mathbf{d}}_q|{\rm e}_q\rangle$.

From the Coulomb-gauge, the electric and magnetic field operators are $\hat{\mathbf{E}}(\mathbf{r},t)=-\partial\hat{\mathbf{A}}(\mathbf{r},t)/\partial t$ and $\hat{\mathbf{B}}(\mathbf{r},t)=\boldsymbol{\nabla}\times\hat{\mathbf{A}}(\mathbf{r},t)$, respectively, where the quantized transverse vector potential is~\cite{Milonni_Book1994}
\begin{align}
\hat{\mathbf{A}}(\mathbf{r},t)=\sum_{\alpha}\sqrt{\frac{\hbar}{2\omega_{\alpha}\varepsilon_0}}\left[\mathbf{u}_{\alpha}(\mathbf{r})\hat{a}_{\alpha}(t)+\mathbf{u}_{\alpha}^*(\mathbf{r})\hat{a}_{\alpha}^{\dagger}(t)\right].
\end{align}
On one hand, note that the quantum properties of the electric and magnetic field operators are determined by the bosonic annihilation and creation operators, $\hat{a}_{\alpha}(t)$ and $\hat{a}_{\alpha}^{\dagger}(t)$, respectively, with usual commutation relations: $[\hat{a}_{\alpha}(t),\hat{a}_{\beta}(t)]=0$ and $[\hat{a}_{\alpha}(t),\hat{a}_{\beta}^{\dagger}(t)]=\delta_{\alpha\beta}$.
On the other hand, the mode functions $\mathbf{u}_{\alpha}(\mathbf{r})$ are classical vector functions satisfying the vector Helmholtz equation and the transversality condition: $[{\nabla}^2+k_{\alpha}^2]\mathbf{u}_{\alpha}(\mathbf{r})=\mathbf{0}$ and $\boldsymbol{\nabla}\cdot\mathbf{u}_{\alpha}(\mathbf{r})=0$, with $k_{\alpha}=\omega_{\alpha}/c$.
These classical functions are chosen to form an orthonormal set: $\int_V {\rm d}^3r\mathbf{u}_{\alpha}^*(\mathbf{r})\cdot\mathbf{u}_{\beta}(\mathbf{r})=\delta_{\alpha\beta}$.
In free space, one has $\mathbf{u}_{\mathbf{k}p}(\mathbf{r})=e^{\imath\mathbf{k}\cdot\mathbf{r}}\mathbf{e}_p/\sqrt{V}$, where $\mathbf{e}_p$ is a polarization vector such that $\mathbf{k}\cdot\mathbf{e}_p=0$ and $V$ is the photon quantization volume.

For weak-coupling between the atoms and the field, one has the contributions to the Hamiltonian~\cite{Milonni_Book1994}: $\hat{\mathcal{H}}_{\rm atom}=\hbar\omega_0\sum_{q=1}^2\hat{S}_q^{z}$, $\hat{\mathcal{H}}_{\rm field}=\sum_{\mathbf{k}p}\hbar\omega_{\mathbf{k}}(\hat{a}_{\mathbf{k}p}^{\dagger}\hat{a}_{\mathbf{k}p}$+1/2), and
\begin{align}
\hat{\mathcal{H}}_{\rm int} = -\imath\hbar\sum_{\mathbf{k}p}\sum_{q=1}^2\left(\hat{S}_q^+ + \hat{S}_q^-\right)\left[g_{\mathbf{k}p}(\mathbf{r}_q)\hat{a}_{\mathbf{k}p}-g_{\mathbf{k}p}^*(\mathbf{r}_q)\hat{a}_{\mathbf{k}p}^{\dagger}\right],
\end{align}
where $\hat{S}_q^{+}=|{\rm e}_q\rangle\langle {\rm g}_q|$ and $\hat{S}_q^-= |{\rm g}_q\rangle\langle {\rm e}_q|$ are the electric dipole raising and lowering operators, respectively, and
\begin{align}
g_{\mathbf{k}p}(\mathbf{r}_q)\equiv\sqrt{\frac{\omega_{\mathbf{k}}}{2\varepsilon_0\hbar}}\mathbf{d}_q\cdot\mathbf{u}_{\mathbf{k}p}(\mathbf{r}_q)\label{g-coupling}
\end{align}
is a complex function associated with the coupling strength between the $q$th atom and field.
The dipole operators satisfy the well-known commutation relations: $[\hat{S}_q^+,\hat{S}_{q'}^-]=2\hat{S}_q^z\delta_{qq'}$, $[\hat{S}_q^z,\hat{S}_{q'}^{\pm}]=\pm\hat{S}_q^{\pm}\delta_{qq'}$, and $[\hat{S}_q^+,\hat{S}_{q'}^-]_+=\delta_{qq'}$, with $(\hat{S}_q^{\pm})^2=0$.

\section{Mode functions for two quantum emitters near a sphere}
\label{Mode-functions}

The influence of the environment on the spontaneous emission rate $\gamma$ is encoded in the classical vector functions $\mathbf{u}_{\mathbf{k}p}$.
For an atom in the vicinity of a spherical body, we can calculate these mode functions from the Lorenz-Mie scattering theory~\cite{Bohren_Book_1983}.
To this end, we have to consider
\begin{align}
\mathbf{u}_{\mathbf{k}p}(\mathbf{r})=\mathbf{u}_{\mathbf{k}p}^{(0)}(\mathbf{r}) + \mathbf{u}_{\mathbf{k}p}^{\rm(s)}(\mathbf{r}),\label{Akp}
\end{align}
where $\mathbf{u}_{\mathbf{k}p}^{(0)}(\mathbf{r})$ is the free space mode function (a plane wave) and $\mathbf{u}_{\mathbf{k}p}^{\rm(s)}(\mathbf{r})$ is the scattering contribution (the returning field) from the spherical body.

Let us set the wave vector $\mathbf{k}=k\mathbf{e}_z$ along the $z$-axis of a coordinate system.
The origin of the coordinate system is located at the center of a sphere of radius $R$ and optical properties $(\varepsilon,\mu)$.
The surrounding medium is the vacuum $(\epsilon_0,\mu_0)$.
We choose a basis of polarization vectors $p=\{\mathbf{e}_x,\mathbf{e}_y\}$ to satisfy $\mathbf{k}\cdot\mathbf{e}_p=0$.
For the polarization along the $x$-axis, and assuming the sphere material is linear, isotropic and non-optically active, one has the expansions in spherical coordinates:
\begin{align}
\mathbf{u}_{\mathbf{k}x}^{\rm(0)}(\mathbf{r})&=\frac{1}{\sqrt{V}}\sum_{\ell=1}^{\infty}A_{\ell}\left[\mathbf{M}_{\ell1}^{(1)}(\mathbf{r})-\imath\mathbf{N}_{\ell1}^{(1)}(\mathbf{r})\right],\label{Ax-inc}\\
\mathbf{u}_{\mathbf{k}x}^{\rm(s)}(\mathbf{r})&=\frac{1}{\sqrt{V}}\sum_{\ell=1}^{\infty}A_{\ell}\left[\imath a_{\ell}\mathbf{N}_{\ell1}^{(3)}(\mathbf{r})-b_{\ell}\mathbf{M}_{\ell1}^{(3)}(\mathbf{r})\right],\label{Ax-sca}
\end{align}
where $A_{\ell} = \imath^{\ell} (2\ell+1)/[\ell(\ell+1)]$, and $\mathbf{M}_{\ell1}$ and $\mathbf{N}_{\ell1}$ are TE (odd) and TM (even) vector spherical harmonic functions~\cite{Bohren_Book_1983}:
\begin{align*}
&\mathbf{M}_{\ell1}(\mathbf{r})=\left[\cos\varphi\pi_{\ell}(\cos\theta)\mathbf{e}_{\theta}-\sin\varphi\tau_{\ell}(\cos\theta)\mathbf{e}_{\varphi}\right]z_{\ell}(kr),\\
&\mathbf{N}_{\ell1}(\mathbf{r})=\cos\varphi\ell(\ell+1)\sin\theta\pi_{\ell}(\cos\theta)\frac{z_{\ell}(kr)}{kr}\mathbf{e}_r\nonumber\\
&+\left[\cos\varphi\tau_{\ell}(\cos\theta)\mathbf{e}_{\theta}-\sin\varphi\pi_{\ell}(\cos\theta)\mathbf{e}_{\varphi}\right]\frac{1}{kr}\frac{{\rm d}[rz_{\ell}(kr)]}{{\rm d}r},
\end{align*}
with $z_\ell(kr)$ being the spherical Bessel function $j_{\ell}(kr)$ for $(1)$ or the Hankel function of first kind $h_{\ell}^{(1)}(kr)$ for $(3)$.
The angle-dependent functions are $\pi_{\ell}(\cos\theta)=P_{\ell}^1(\cos\theta)/\sin\theta$ and $\tau_{\ell}(\cos\theta)={\rm d}P_{\ell}^1(\cos\theta)/{\rm d}\theta$, where $P_{\ell}^1(\cos\theta)$ is the associated Legendre function of first order.
The vector mode functions $\mathbf{u}_{\mathbf{k}y}^{\rm(0)}(\mathbf{r})$ and $\mathbf{u}_{\mathbf{k}y}^{\rm(s)}(\mathbf{r})$ for the polarization along the $y$-axis can be readily obtained by changing $\varphi\to\varphi+\pi/2$ in Eqs.~(\ref{Ax-inc}) and (\ref{Ax-sca}), respectively.
In particular, the Lorenz-Mie coefficients $a_{\ell}$ and $b_{\ell}$ are determined by boundary conditions; for the simplest case of a center-symmetric core-shell sphere, with inner radius $a$ and outer radius $b$, they read
\begin{align}
        a_{\ell} &=\frac{\widetilde{n}_{\rm s}\psi_{\ell}'(kb)-\psi_{\ell}(kb)\mathcal{A}_{\ell}(n_{\rm s}kb)}{\widetilde{n}_{\rm s}\xi_{\ell}'(kb)-\xi_{\ell}(kb)\mathcal{A}_{\ell}(n_{\rm s}kb)},\label{an}\\
        b_{\ell} &=\frac{\psi_{\ell}'(kb)-\widetilde{n}_{\rm s}\psi_{\ell}(kb)\mathcal{B}_{\ell}(n_{\rm s}kb)}{\xi_{\ell}'(kb)-\widetilde{n}_{\rm s}\xi_{\ell}(kb)\mathcal{B}_{\ell}(n_{\rm s}kb)},\label{bn}
\end{align}
with the auxiliary functions being
\begin{align*}
\mathcal{A}_{\ell}(n_{\rm s}kb)&=\frac{\psi_{\ell}'(n_{\rm s}kb)-\widetilde{A}_{\ell}\chi_{\ell}'(n_{\rm s}kb)}{\psi_{\ell}(n_{\rm s}kb)-\widetilde{A}_{\ell}\chi_{\ell}(n_{\rm s}kb)},\\
\mathcal{B}_{\ell}(n_{\rm s}kb)&=\frac{\psi_{\ell}'(n_{\rm s}kb)-\widetilde{B}_{\ell}\chi_{\ell}'(n_{\rm s}kb)}{\psi_{\ell}(n_{\rm s}kb)-\widetilde{B}_{\ell}\chi_{\ell}(n_{\rm s}kb)},\\
        \widetilde{A}_{\ell} &= \frac{\widetilde{n}_{\rm s}\psi_{\ell}(n_{\rm s}ka)\psi_{\ell}'(n_{\rm c}ka)-\widetilde{n}_{\rm c}\psi_{\ell}'(n_{\rm s}ka)\psi_{\ell}(n_{\rm c}ka)}{\widetilde{n}_{\rm s}\chi_{\ell}(n_{\rm s}ka)\psi_{\ell}'(n_{\rm c}ka)-\widetilde{n}_{\rm c}\chi_{\ell}'(n_{\rm s}ka)\psi_{\ell}(n_{\rm c}ka)},\\
        \widetilde{B}_{\ell}&=\frac{\widetilde{n}_{\rm s}\psi_{\ell}'(n_{\rm s}ka)\psi_{\ell}(n_{\rm c}ka)-\widetilde{n}_{\rm c}\psi_{\ell}(n_{\rm s}ka)\psi_{\ell}'(n_{\rm c}ka)}{\widetilde{n}_{\rm s}\chi_{\ell}'(n_{\rm s}ka)\psi_{\ell}(n_{\rm c}ka)-\widetilde{n}_{\rm c}\chi_{\ell}(n_{\rm s}ka)\psi_{\ell}'(n_{\rm c}ka)},
\end{align*}
where the functions $\psi_{\ell}(z)=z j_{\ell}(z)$, $\chi_{\ell}(z)=-z y_{\ell}(z)$ and $\xi_{\ell}(z)=\psi_{\ell}(z)-\imath\chi_{\ell}(z)$ are the Riccati-Bessel, Riccati-Neumann and Riccati-Hankel functions, respectively, with $j_{\ell}$ and $y_{\ell}$ being the spherical Bessel and Neumann functions~\cite{Bohren_Book_1983}.
The relative refractive and impedance indices (in relation to the surrounding medium) are $n_q=\sqrt{\varepsilon_{q}\mu_{q}/(\varepsilon_0\mu_0)}$ and $\widetilde{n}_q = \sqrt{\varepsilon_{q}\mu_0/(\varepsilon_0\mu_{q})}$, with $q={\rm c}$ for the core and $q={\rm s}$ for the shell~\cite{Arruda_JOSA27_1_2010,Arruda_PhysRevA87_2013}.
For nonmagnetic materials ($\mu_q=\mu_0$), one has $\widetilde{n}_q=n_q$~\cite{Arruda_JOpt14_2012}.

To calculate the spontaneous emission rate of photons using Eq.~(\ref{Gamma-single}) and (\ref{Gamma-coupling}), note that we have to integrate over $k$-space rather than position. Indeed, in order to calculate Eqs.~(\ref{Gamma-single}) and (\ref{Gamma-coupling}),
we have to make the well-known replacement:
\begin{align}
\sum_{\mathbf{k}p} \longrightarrow\lim_{V\to\infty}\sum_{p}\frac{V}{8\pi^3}\int{\rm d}^3k.
\end{align}
Usually, this change of coordinates from $(r,\theta,\varphi)$ to $(k,\theta_k,\varphi_k)$ is not straightforward and requires the use of the addition theorem of vector spherical harmonics, which is widely used in multiple scattering schemes~\cite{Ng2005}.
However, as discussed in Ref.~\cite{Farina_PhysRevA87_2013}, by simple geometric arguments one can verify that the mode functions in $k$-space are obtained from Eqs.~(\ref{Ax-inc}) and (\ref{Ax-sca}) by changing $k\to k$, $\theta\to-\theta_k$ and $\varphi\to0$, where now we have to consider $\mathbf{e}_k$ instead of $\mathbf{e}_r$.
By this procedure, we finally arrive at
\begin{align}
\mathbf{u}_{\mathbf{k}x}^{\rm(0)}(\mathbf{r})&=\frac{1}{\sqrt{V}}\sum_{\ell=1}^{\infty}\frac{A_{\ell}}{kr}\big\{\imath\sin\theta_k j_{\ell}(kr)\ell(\ell+1)\pi_{\ell}\mathbf{e}_k\nonumber\\
&+\left[\pi_{\ell}\psi_{\ell}(kr)-\imath\tau_{\ell}\psi_{\ell}'(kr)\right]\mathbf{e}_{\theta_k}\big\},\label{Akx-inc}\\
\mathbf{u}_{\mathbf{k}x}^{\rm(s)}(\mathbf{r})&=\frac{1}{\sqrt{V}}\sum_{\ell=1}^{\infty}\frac{A_{\ell}}{kr}\big\{-\imath\sin\theta_k a_{\ell} h_{\ell}^{(1)}(kr)\ell(\ell+1)\pi_{\ell}{\mathbf{e}_k}\nonumber\\
&-\left[b_{\ell}\pi_{\ell}\xi_{\ell}(kr)-\imath a_{\ell}\tau_{\ell}\xi_{\ell}'(kr)\right]\mathbf{e}_{\theta_k}\big\},\\
\mathbf{u}_{\mathbf{k}y}^{\rm(0)}(\mathbf{r})&=\frac{1}{\sqrt{V}}\sum_{\ell=1}^{\infty}\frac{A_{\ell}}{kr}\left[\imath\pi_{\ell}\psi_{\ell}'(kr)-\tau_{\ell}\psi_{\ell}(kr)\right]\mathbf{e}_{\varphi_k},\label{Aky-inc}\\
\mathbf{u}_{\mathbf{k}y}^{\rm(s)}(\mathbf{r})&=\frac{1}{\sqrt{V}}\sum_{\ell=1}^{\infty}\frac{A_{\ell}}{kr}\left[-\imath a_{\ell}\pi_{\ell}\xi_{\ell}'(kr)+b_{\ell}\tau_{\ell}\xi_{\ell}(kr)\right]\mathbf{e}_{\varphi_k}.\label{Aky-sca}
\end{align}

\section{Local fields in the vicinity of a sphere}
\label{Lorenz-Mie}

Based on the Lorenz-Mie theory, we consider the local field in the vicinity of a sphere as a classical electromagnetic wave with time-harmonic dependence $e^{-\imath\omega t}$.
The electric field consists of an incoming plane wave propagating along the positive $z$ direction and the corresponding scattered field by the sphere~\cite{Bohren_Book_1983}:
\begin{align}
\mathbf{E}(\mathbf{r}) = \mathbf{E}_{\rm in}(\mathbf{r}) + \mathbf{E}_{\rm sca}(\mathbf{r}),\label{E-laser}
\end{align}
where the incoming electric field is $\mathbf{E}_{\rm in}(\mathbf{r})=E_0\mathbf{e}_x e^{\imath\mathbf{k}\cdot\mathbf{r}}$ and
\begin{align*}
\mathbf{E}_{\rm sca}(\mathbf{r})&=\frac{1}{kr}\sum_{\ell=1}^{\infty}E_{\ell}\bigg\{\imath\cos\varphi\sin\theta a_{\ell} h_{\ell}^{(1)}(kr)\ell(\ell+1)\pi_{\ell}\mathbf{e}_{{r}}\nonumber\\
&-\cos\varphi\left[b_{\ell}\pi_{\ell}\xi_{\ell}(kr)-\imath a_{\ell}\tau_{\ell}\xi_{\ell}'(kr)\right]\mathbf{e}_{{\theta}}\nonumber\\
&-\sin\varphi\left[\imath a_{\ell}\pi_{\ell}\xi_{\ell}'(kr)-b_{\ell}\tau_{\ell}\xi_{\ell}(kr)\right]\mathbf{e}_{\varphi}\bigg\}
\end{align*}
is the scattered electric field, with $E_{\ell} = \imath^{\ell} E_0(2\ell+1)/[\ell(\ell+1)]$ and $\mathbf{k}=k{\mathbf{e}}_z$.
The polarization along $y$ direction is obtained by replacing $(\mathbf{e}_x;\varphi)$ with $(\mathbf{e}_y;\varphi+\pi/2)$ in the expressions of $\mathbf{E}_{\rm in}$ and $\mathbf{E}_{\rm sca}$.
The Lorenz-Mie coefficients $a_{\ell}$ and $b_{\ell}$ are evaluated at the angular frequency $\omega$.
In the far field $(kr\gg1)$, one has ${E}_r\approx0$ and $E_{\theta},E_{\varphi}\propto E_0(e^{\imath kz}+e^{\imath kr}/kr)$.

\end{document}